\documentclass[12pt,a4paper]{article}                           

\usepackage{amsmath}
\usepackage{amssymb}
\usepackage[usenames]{color} 
\usepackage{multirow}
\usepackage{graphicx}
\usepackage{epstopdf}
\usepackage{array}
\usepackage{hhline}
\usepackage{cite}
\usepackage{microtype}
\usepackage[bf]{caption}
\usepackage{color}
\usepackage[usenames,dvipsnames]{xcolor}
\usepackage{subcaption}
\usepackage{pstool}
\usepackage{bbold}
\usepackage{mathtools}
\usepackage{todonotes}

\usepackage{verbatim}

\usepackage[hang]{footmisc}
\renewcommand{\footnotesize}{\small}

\usepackage{ifpdf}
\ifpdf
  \usepackage[pdftex,
    pdftitle={},
    pdfauthor={},
    pdfsubject={},
    bookmarksopen, bookmarksnumbered, bookmarksopenlevel=2]{hyperref}
\fi
\def\hybrid{\topmargin -20pt    \oddsidemargin 0pt
        \headheight 0pt \headsep 0pt
        \textwidth 6.25in       
        \textheight 9 in       
        \marginparwidth .875in
        \parskip 5pt plus 1pt 
          \jot = 1.5ex
   }
\hybrid
\numberwithin{equation}{section}
\numberwithin{table}{section}

\newcommand{\beq}{\begin{equation}}  \newcommand{\eeq}{\end{equation}}
\newcommand{\bal}{\begin{aligned}}   \newcommand{\eal}{\end{aligned}}
\newcommand{\bea}{\begin{eqnarray}}  \newcommand{\eea}{\end{eqnarray}}

\newcommand{\bmat}{\left(\begin{array}}
\newcommand{\emat}{\end{array}\right)}

\newcommand{\nn}{\nonumber}

\newcommand{\cO}{\mathcal{O}}

\newcommand{\cC}{\mathcal{C}}

\newcommand{\cL}{\mathcal{L}}

\newcommand{\cK}{\mathcal{K}}
\newcommand{\cN}{\mathcal{N}}
\newcommand{\cW}{\mathcal{W}}

\newcommand{\cA}{\mathcal{A}}

\newcommand{\cV}{\mathcal{V}}

\newcommand{\R}{\text{Re}}

\usepackage{cancel}

\newcommand{\be}{\begin{equation}}
\newcommand{\ee}{\end{equation}}

\usepackage{amsthm}
\usepackage[framemethod=TikZ]{mdframed}
\mdfsetup{skipabove=\topskip, skipbelow=\topskip}

\def\K{\mathcal{K}}
\def\W{\mathcal{W}}

\newcommand{\kom}{\, ,\quad }
\newcommand*{\raw}{\rightarrow}

\newcommand*{\p}{\mathop{}\!\mathrm \partial}

\begin{document}

\baselineskip=14pt
\parskip 5pt plus 1pt

\vspace*{-1.5cm}
\begin{flushright}    
  {\small 
  
  }
\end{flushright}

\vspace{2cm}
\begin{center}        
  {\LARGE Large Field Ranges from Aligned\\[0.5cm]
  and Misaligned Winding}
\end{center}

\vspace{0.5cm}
\begin{center}        
{\large  Arthur Hebecker, Daniel Junghans and Andreas Schachner}
\end{center}

\vspace{0.15cm}
\begin{center}        
  \emph{Institut f\"ur Theoretische Physik, Ruprecht-Karls-Universit\"at, \\
             Philosophenweg 19, 69120 
             Heidelberg, Germany} \\[0.15cm]
             \vspace{1cm}
(\texttt{\href{mailto:a.hebecker@thphys.uni-heidelberg.de}{a.hebecker@thphys.uni-heidelberg.de}, \href{mailto:junghans@thphys.uni-heidelberg.de}{junghans@thphys.uni-heidelberg.de}, \href{mailto:schachner@thphys.uni-heidelberg.de}{schachner@thphys.uni-heidelberg.de}}) \\[0.15cm]
\vspace{1cm}
 21.12.2018 

\end{center}

\vspace{1cm}


\begin{abstract}
\noindent
We search for effective axions with super-Planckian decay constants in type IIB string models. We argue that such axions can be realised as long winding trajectories in complex-structure moduli space by an appropriate flux choice. Our main findings are: The simplest models with {\it aligned} winding in a 2-axion field space fail due to a general no-go theorem. However, equally simple models with {\it misaligned} winding, where the effective axion is not close to any of the fundamental axions, appear to work to the best of our present understanding. These models have large decay constants but no large monotonic regions in the potential, making them unsuitable for large-field inflation.
We also show that our no-go theorem can be avoided by aligning three or more axions. We argue that, contrary to misaligned models, such models can have both large decay constants and large monotonic regions in the potential. Our results may be used to argue against the refined Swampland Distance Conjecture and strong forms of the axionic Weak Gravity Conjecture.
It becomes apparent, however, that realising inflation is by far harder than just producing a light field with large periodicity.
\end{abstract}

\thispagestyle{empty}
\clearpage



\newpage

\tableofcontents

\section{Introduction}

One of the most prominent aspects of the landscape-swampland program \cite{Vafa:2005ui,Ooguri:2006in,ArkaniHamed:2006dz} is the quest for large field ranges in string compactifications. One reason for this is the interest in large-field inflation. Another is the hope for a deeper understanding of general quantum gravity constraints and therefore of quantum gravity itself.

In the present paper, we focus on large axionic field ranges. We do not take the road of monodromy \cite{Silverstein:2008sg,McAllister:2008hb} or its modern variant of $F$-term axion monodromy \cite{Marchesano:2014mla,Blumenhagen:2014gta,Hebecker:2014eua}. Instead, we pursue the idea of constructing an effective large-$f$ axion starting from two or more fundamental axions in the UV \cite{Kim:2004rp}.
Specifically, we argue that such constructions can plausibly be realised using flux constraints in the complex-structure sector of type IIB string theory~\cite{Hebecker:2015rya}.
The main limitation is that we do not (yet) have an explicit geometry and a concrete flux choice. If our results stand up, they arguably lead to a tension with the Weak Gravity Conjecture (WGC) \cite{ArkaniHamed:2006dz}, at least in some of its stronger forms (for recent analyses in the axion context, see, e.g., \cite{Rudelius:2014wla, Rudelius:2015xta, delaFuente:2014aca, Montero:2015ofa, Brown:2015iha, Brown:2015lia, Bachlechner:2015qja, Hebecker:2015rya, Junghans:2015hba, Heidenreich:2015wga, Heidenreich:2015nta, Heidenreich:2016aqi, Ibanez:2015fcv, Kooner:2015rza,Kappl:2015esy, Hebecker:2015zss, Hebecker:2016dsw, Hebecker:2017wsu, Hebecker:2017uix, Palti:2017elp, Lust:2017wrl}). In addition, our effective axion with parametrically large $f$ might be interpreted as violating the refined form \cite{Klaewer:2016kiy} of the Swampland Distance Conjecture (SDC) \cite{Vafa:2005ui,Ooguri:2006in}.

Before discussing our concrete setup, let us qualify what we mean by a parametrically large field range: There are many examples in string theory of infinite directions in field space. However, in all such known examples, moving super-Planckian distances causes a tower of states to become exponentially light \cite{Ooguri:2006in, Klaewer:2016kiy, Baume:2016psm, Valenzuela:2016yny, Blumenhagen:2017cxt, Palti:2017elp, Cicoli:2018tcq, Grimm:2018ohb, Heidenreich:2018kpg, Blumenhagen:2018nts, Blumenhagen:2018hsh, Dias:2018pgj, Grimm:2018cpv} (see \cite{Hebecker:2017lxm, Landete:2018kqf} for caveats). This implies an exponentially falling cut-off. By parametrically large field distance we mean a distance $\sim N\cdot M_p$, with $N\gg 1$ a flux number, over which no such light tower appears. In this sense, our constructions might serve as counter-examples to the refined SDC,
possibly calling for a weakening of the claim.
We find this interesting {\it independently} of whether the potential of the emerging large-$f$ axion turns out to be suitable for inflation. Indeed, it will become clear that obtaining a large-$f$ effective axion {\it unsuitable} for inflation is the simpler task. To turn this into a model of natural inflation, one must avoid short-range oscillations in the axion potential and stabilise moduli at a fairly high scale. This is much more demanding.

Our basic method is the restriction of a multi-axion field space to a winding trajectory by an appropriate flux choice~\cite{Hebecker:2015rya}.\footnote{This can be viewed as the Higgsing of several 0-forms by $(-1)$-forms \cite{Dvali:2005an,Kaloper:2008fb,Kaloper:2011jz}, such that a single 0-form with large $f$ survives. Similarly, several $1$-forms can be Higgsed by 0-forms to challenge the WGC for vector fields \cite{Saraswat:2016eaz}. Thus, establishing the original proposal of \cite{Hebecker:2015rya} would be important to evaluate how much trust one can put in the subsequent more general claim of \cite{Saraswat:2016eaz}. 
}
Concretely, certain linear combinations of complex-structure axions receive a mass from type IIB 3-form fluxes such that only a one-dimensional, potentially very long, winding trajectory survives. In the large-complex-structure limit and at tree-level, the corresponding axion is exponentially light. Originally, this was suggested as a model of `winding inflation'~\cite{Hebecker:2015rya}, see also \cite{Kobayashi:2015aaa,Bizet:2016paj,Blumenhagen:2016bfp,Wolf:2017wmu}.\footnote{Shift-symmetric complex-structure moduli have been considered in the context of inflation before, e.g., as complex-structure moduli of 4-folds or D7-brane moduli~\cite{Hebecker:2011hk,35,Arends:2014qca,Ibanez:2014swa,Carta:2016ynn,Landete:2017amp,Kim:2018vgz} as well as in the 3-fold case~\cite{Garcia-Etxebarria:2014wla,Abe:2014xja}.}
Subsequently, it was pointed out that, in related type IIA models, a parametrically large $f$ strongly constrains the achievable instanton hierarchy and hence the potential~\cite{Palti:2015xra,Baume:2016psm}. In particular, it was argued there that a large-$f$ effective axion can be constructed in the mirror-dual of $\mathbb{CP}^4_{(1,1,2,2,6)}[12]$ but no monotonic region suitable for inflation exists. In fact, the situation is complicated further in this model because, as we will show, flux-backreaction becomes a troubling factor. While it is unclear whether these issues are generic in type IIA, we will argue that they can be avoided in type IIB.\footnote{See, however, \cite{Blumenhagen:2016bfp,Wolf:2017wmu} for a critical discussion of large field ranges in type IIB models at the conifold point. For very recent optimistic analyses in a rather different approach see \cite{Hebecker:2018yxs, Buratti:2018xjt}.}
Furthermore, in comparison to type IIB, type IIA constructions
do not allow for an easy separation between the masses of the complex-structure moduli and the AdS scale, and less is known about possible uplifting mechanisms. It is hence mandatory to understand the type IIB situation.

\begin{figure}[t!]
\centering
 \includegraphics[scale=0.30]{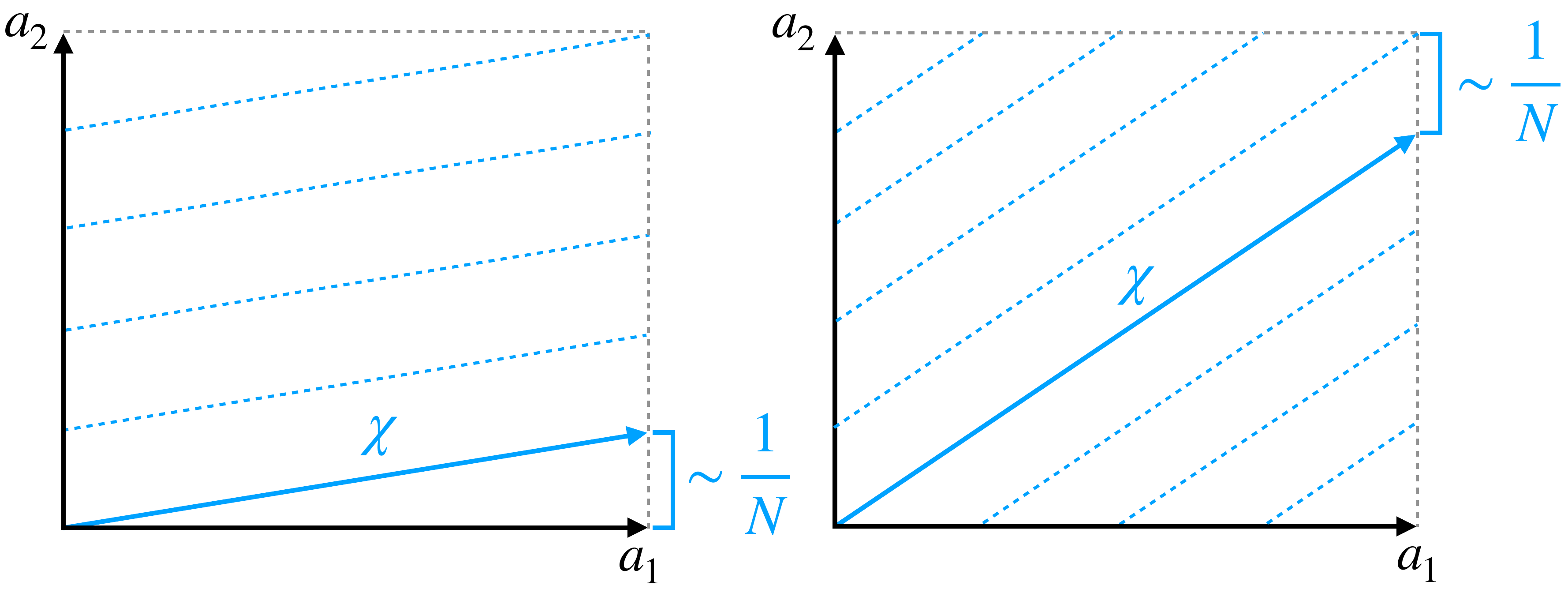}
\caption{Aligned winding (left) and misaligned winding (right) in a two-dimensional axion field space. The blue arrow corresponds to the light axion direction $\chi$, while the axes represent the two fundamental axions $a_1$ and $a_2$. The light axion is either almost aligned with a fundamental axion or with the diagonal, where the angle is controlled by the parameter $1/N$ such that perfect (mis-)alignment corresponds to $N\raw\infty$.}\label{fig:almal}
\end{figure}

A reasonable strategy is to first establish examples of large field ranges in type IIB before addressing the even more difficult task of large-field inflation.
Recently, it was found in \cite{Hebecker:2017lxm} that large field ranges can indeed be obtained in simple toroidal models. This raises hopes that one can actually construct low-energy effective field theories (EFTs) for axions with parametrically large effective decay constants as part of the landscape. By this we mean working with general Calabi-Yau threefolds and stabilising the saxions. The purpose of this paper is to perform a detailed analysis of this possibility.

We first study the simplest case of an aligned winding trajectory with two fundamental axions (cf.~the left-hand side of Fig.~\ref{fig:almal}). We find a general no-go theorem for this scenario, stating that parametrically large field ranges are ruled out on any Calabi-Yau orientifold. We then propose several variants of the winding idea which allow to avoid the no-go theorem and may thus lead to large field ranges. First, we study 'misaligned winding', where we consider a light axion direction aligned with a diagonal of the axion field space rather than one of its axes (cf.~the right-hand side of Fig.~\ref{fig:almal}). Second, we consider constructions with a finely tuned superpotential. Third, we consider aligned winding of three or more fundamental axions. We also analyse the prospects for aligned or misaligned winding in the concrete setting of the large-complex-structure (LCS) limit, where the $F$-term constraints are simple enough to be solved in complete generality. Interestingly, we find that the problem of constructing long winding trajectories in this setting can be reduced to a purely geometric condition involving the triple intersection numbers of the Calabi-Yau.

The paper is organised as follows. In Sect.~\ref{sec:fdwgc}, we expand on the relation between axion field distances and the WGC and discuss the mechanisms of aligned and misaligned winding. We furthermore discuss winding in the context of a simple type IIA compactification and discuss some problems that occur in this model. In Sect.~\ref{sec:fluxwiniib}, we study axion field ranges in the complex-structure sector of type IIB Calabi-Yau compactifications. We first establish a no-go theorem for aligned winding with two axions and then study several approaches that avoid this result. In Sect.~\ref{sec:EFT_LF_AXION}, we derive the low-energy EFT for the light axion with parametrically large $f$. In particular, we discuss Kahler moduli stabilisation and show that there is a regime where a hierarchy between all moduli masses and our large-$f$ effective axion potential is guaranteed. We furthermore discuss the challenges that arise when promoting our scenario to a model of large-field inflation. We summarise our results in Sect.~\ref{sec:summary}.

\section{General idea}
\label{sec:fdwgc}

\subsection{(Mis-)aligned winding and swampland conjectures}
\label{sec:amis}

Motivated by understanding field distances for fields with potentials, we are naturally led to looking at axions. This is because their potential is well-controlled due to a discrete shift symmetry. The axion version of the Weak Gravity Conjecture reads \cite{ArkaniHamed:2006dz}
\be
f S \lesssim q \,. \label{wgc1a}
\ee
Here, and henceforth, we set the Planck mass to unity, $M_p=1$. Furthermore, $f$ is the axion decay constant, $S$ is the action of the instanton satisfying the inequality and $q$ is its charge. The action for a canonically normalised axion is periodic under a shift $2\pi f$. Let us furthermore denote by $2\pi \Pi$ the periodicity of the potential generated by the instanton. We then have
\be
\Pi = \frac{f}{q} \,. \label{Piq}
\ee
It is important to note that this can be different from the periodicity of the axion since it is possible to have $q \gg 1$.

Depending on the instanton(s) satisfying \eqref{wgc1a}, one can distinguish different versions of the WGC. In particular, the Strong WGC \cite{ArkaniHamed:2006dz} states that (\ref{wgc1a}) is satisfied for the instanton with the smallest action $S$. In the controlled instanton regime $S\gtrsim 1$, \eqref{wgc1a} implies $f \lesssim q$ and hence $\Pi \lesssim 1$. Therefore, if the Strong WGC holds, the axion potential is dominated by an instanton contribution with sub-Planckian periodicity, ruling out, e.g., large-field inflation.\footnote{This is true modulo the small-action loophole pointed out in \cite{delaFuente:2014aca} (see also \cite{Hebecker:2018ofv}).} Conversely, an axion with a parametrically large monotonic region in the potential parametrically violates the Strong WGC. Note, however, that the Strong WGC does not impose any restriction on $f$, and hence, for large enough $q$, the axion field range can still be super-Planckian.

This is to be contrasted with the Smallest Charge WGC \cite{ArkaniHamed:2006dz}, which states that (\ref{wgc1a}) is satisfied for an instanton with $q=1$. In the regime $S\gtrsim 1$, this implies $f \lesssim 1$ and hence a small field range. The Smallest Charge WGC is thus much more restrictive than the Strong WGC.\footnote{Another reason why, despite its name, the Strong WGC is less strong than the Smallest Charge WGC is that its 1-form version does not have any implications for the spectrum of the low-energy EFT. In particular, if only the Strong WGC holds, the inequality $m\lesssim qg$ \cite{ArkaniHamed:2006dz} can be satisfied by states with arbitrarily large charges and, hence, arbitrarily large masses.}
A variant of the Smallest Charge WGC is the (Sub-)Lattice WGC \cite{Heidenreich:2016aqi, Montero:2016tif}, where an instanton satisfying \eqref{wgc1a} exists for every site on the charge lattice or, more generally, on a sub-lattice with coarseness $q_c\ge 1$.\footnote{See also the less restrictive Tower WGC \cite{Andriolo:2018lvp}, where the WGC is also satisfied by a large number of states but they do not necessarily occupy a sub-lattice in charge space.}
While an apparent counter-example \cite{ArkaniHamed:2006dz} to the case $q_c=1$ was later shown to be incorrect \cite{Heidenreich:2015nta}, there are more recent counter-examples which indeed violate the Smallest Charge/Lattice WGC for $q_c=1$ \cite{Heidenreich:2016aqi, Montero:2016tif}. We stress, however, that it is an open problem whether the Sub-Lattice WGC is true for sub-lattices with $q_c>1$ (but still $\mathcal{O}(1)$). As we will see below, our type IIB constructions in Sect.~\ref{sec:fluxwiniib} correspond to a parametric violation of this statement, i.e., to sub-lattices with parametrically large $q_c$.
We will furthermore argue in Sect.~\ref{sec:fluxwiniib} that it may be possible to construct axion potentials with large monotonic regions in our type IIB setting. According to our above discussion, this would correspond to a parametric violation of the Strong WGC.

We would like to test axion field distances and the related versions of the WGC directly in string theory. We are therefore interested in setups which lead to $f \gg 1$. The most reliable settings utilise constructions with two or more axions, in the spirit of \cite{Kim:2004rp} (see also \cite{Berg:2009tg,Ben-Dayan:2014zsa}). The idea is to give all but one combination of them a large mass and study the remaining light direction in the resulting effective theory. The key property of a setup with several axions is the possibility of winding trajectories for the light direction by turning on fluxes \cite{Hebecker:2015rya}. In the simple case of two axions, this winding can be achieved by considering a superpotential of the form
\begin{equation}\label{eq:W1} 
\W=w(Z)+f(Z)(n_{1}\, U_{1}+n_{2}\, U_{2})\,.
\end{equation}
Here, $n_{1},n_{2}\in\mathbb{Z}$ are fluxes and $U_{1},U_{2}$ correspond to two distinguished complex-structure moduli. The remaining moduli are denoted by $Z$. The axions, $a_1$ and $a_2$, are the real parts of the $U_i$, while the saxions, $u_1$ and $u_2$, are the imaginary parts: $U_i = a_i+iu_i$. The associated decay constants are denoted by $f_1$ and $f_2$, where we assume $f_1,f_2 \lesssim 1$. The axions have associated instantons with actions $S_1=u_1$, $S_2=u_2$. Assuming that the instantons are unit-charged, we furthermore have $\Pi_1=f_1$ and $\Pi_2=f_2$.

The combination of axions which obtains a large mass from \eqref{eq:W1} is $n_1 a_1 + n_2 a_2$. We are interested in the effective theory of the surviving light axion combination. To quantify the effective field range, it is useful to introduce the co-prime parts of $n_1$ and $n_2$, so we write
\be
n_1 = p \,p_1 \,,\quad n_2 = p \,p_2 \,,\qquad \mathrm{for\;largest\;} p \in \mathbb{Z}\; \mathrm{such\;that\;}p_1,p_2 \in \mathbb{Z}\,.
\label{pdef}
\ee
The potential is invariant under any axion shift $(\Delta a_1, \Delta a_2)$ orthogonal to $(p_1, p_2)$. Hence, we can parametrise the flat direction by some field $\chi$ as
\begin{equation}\label{eq:Def:Chi} 
\left (\begin{array}{c}
a_{1} \\ 
a_{2}
\end{array} \right )=\left (\begin{array}{c}
-p_{2} \\ 
p_{1}
\end{array}\right )\, \chi\; .
\end{equation}
Here, without loss of generality, we redefined the fundamental axions $a_1$, $a_2$ such that the line parametrised by $\chi$ goes through $(0,0)$.
The vector on the right-hand side is the smallest integer vector pointing along the flat direction so that $\chi$ is $2\pi$-periodic. The canonically normalised field obtained from $\chi$ will be denoted by $\psi$. At leading order in the mass ratios of the heavy and light axion combinations, we can extract the effective decay constant for the light direction $\psi$ by treating the massive axion as constant in the kinetic terms. In the absence of kinetic mixing between $a_1$ and $a_2$, we then find
\begin{align}
{\cal L} &= f_1^2 \left(\partial a_1 \right)^2 + f_2^2 \left(\partial a_2 \right)^2+A\, \mathrm{e}^{-S_{1}}\,\cos(a_{1})+B\, \mathrm{e}^{-S_{2}}\,\cos(a_{2}) \nn\\
&\simeq  \left( p_2^2 f_1^2 + p_1^2 f_2^2 \right) \left(\partial \chi \right)^2 +A\, \mathrm{e}^{-S_{1}}\,\cos\left (p_{2}\, \chi \right )+B\, \mathrm{e}^{-S_{2}}\,\cos\left (p_{1}\, \chi \right )\nn\\
&\equiv \left(\partial \psi \right)^2+A\, \mathrm{e}^{-S_{1}}\,\cos\left (\dfrac{\psi}{\Pi_{1}}\right )+B\, \mathrm{e}^{-S_{2}}\,\cos\left (\dfrac{\psi}{\Pi_{2}}\right )\,. \label{P2in2}
\end{align}
Note that, in the first line, we only displayed the instanton-generated part of the potential for $a_1$, $a_2$ and assumed for simplicity that there are no relative phases in the arguments of the cosines.
The periodicities of $\psi$ in the two instantons, associated to $a_1$ and $a_2$, are
\be
\Pi_1 = \frac{f}{p_2} \,,\quad \Pi_2 = \frac{f}{p_1} \,,
\label{P2in}
\ee
where
\be
f = \left( p_1^2 f_2^2 + p_2^2 f_1^2 \right)^{\frac12} \,.
\label{fmixp}
\ee
The field range is determined by the periodicity of the full action. This is equivalent to the periodicity under the two instanton terms appearing in \eqref{P2in2}, i.e., both terms must be periodic under a single axion shift. Since $p_1$ and $p_2$ are co-prime, the periodicity of $\psi$ is then $f$, and this sets the field range. Comparing with (\ref{Piq}), we also see that the associated charges of the instantons are $p_1$ and $p_2$.

We can now distinguish, within this setting, two scenarios for obtaining a large field range for $\psi$. First, we consider \emph{aligned winding}, i.e.,
\be \label{al}
p_1 \sim {\cal O}\left(1\right) \,,\quad p_2 \sim {\cal O}\left(N\right) \,,
\ee
for some $N \gg 1$. This is illustrated as the first case in Fig.\ \ref{fig:almal}. According to \eqref{P2in} and \eqref{fmixp}, it allows for $f \gg 1$ as well as $\Pi_2 \gg 1$. The second condition may admit large monotonic regions in the potential and possibly even inflation. For this reason, the alignment scenario was originally proposed in \cite{Kim:2004rp}.  As dicussed above, this implies a parametric violation of the Smallest Charge WGC and the Strong WGC. If the latter holds, the sub-Planckian instanton, with periodicity $\Pi_1$, dominates the super-Planckian one (i.e., $S_2 > S_1$) such that the monotonic regions in the potential are small.

The other scenario we consider is \emph{misaligned winding} where we take
\be \label{mal}
p_1 \sim {\cal O}\left(N\right) \,,\quad p_2 \sim {\cal O}\left(N\right)
\ee
with $p_1 \neq p_2$. This is illustrated as the second case in Fig.\ \ref{fig:almal}.
Here, it is manifest that the periodicities in the instantons are not parametrically large (cf.\ \eqref{P2in} and \eqref{fmixp}), and so it would not be useful for inflation. However, the axion field range $f$ is parametrically enhanced such that we still have $f \gg 1$. We will exploit both aligned and misaligned winding to realise large field ranges in type IIB string theory.\footnote{More generally, one could also consider flux choices such as $p_1\sim\mathcal{O}(N)$, $p_2\sim\mathcal{O}(N^2)$, which share some of the features of both aligned and misaligned winding. In particular, the present example allows $\Pi_2\gg 1$ as in aligned winding and yields $f \gg \Pi_1,\Pi_2$ as in misaligned winding. We will not consider such flux configurations in the remainder of this paper as the cases \eqref{al} and \eqref{mal} are sufficient for the points we wish to make.}

Let us see how misaligned winding fits into the framework of the WGC. We assume that there are two instantons associated to $a_1$ and $a_2$ which satisfy the WGC, i.e., $S_i f_i \lesssim 1$.
For simplicity, we furthermore take $f_1 = f_2$ and $p_1 \sim p_2 \sim N$. It then follows with \eqref{fmixp} that
\be
f S_i \lesssim N \,,
\ee
i.e., the WGC is satisfied by instantons with parametrically large charges under the light axion $\psi$. To consider the low-charge instantons, we would need to take instantons wrapping cycles in a homology class which is a linear combination of the homology classes of the 'fundamental' cycles associated to $u_1$ and $u_2$. But the instantons we considered are the lightest leading instantons, and so the instantons with lower charges have a larger action. For $f\sim\mathcal{O}(N)$, they will therefore not satisfy the WGC inequality. The setup therefore amounts to a parametric violation of the Smallest Charge WGC \cite{ArkaniHamed:2006dz}. Within the context of the Sub-Lattice WGC \cite{Heidenreich:2016aqi, Montero:2016tif}, it amounts to a sub-lattice with parametrically large coarseness, as discussed above.

Let us finally address the Swampland Distance Conjecture. Applied to the case of our axion $\chi$, the refined SDC states that a tower of states with exponentially light masses $m \sim \mathrm{e}^{-c\Delta\chi}$, $c>0$ should appear as we move a super-Planckian distance $\Delta\chi\gtrsim 1$ in the axion field direction. However, the energy scale of the axion potential we constructed is exponentially small compared to the mass scales of the other moduli. This means that the variation of $\chi$ cannot generate a large backreaction and, consequently, the (moduli-dependent) masses of any tower of states will stay approximately constant. We therefore conclude that no tower of light states can appear as we move along the axion potential, i.e., the above construction, if realisable in string theory, could provide a counter-example to the refined SDC.

\subsection{A type IIA example}
\label{sec:axalCY}

We now consider the winding scenario in AdS solutions of type IIA string theory compactified on a Calabi-Yau orientifold \cite{DeWolfe:2005uu}.\footnote{These solutions are only known in the smeared limit, i.e., they do not take into account the local backreaction of the O6-planes. We will proceed with the assumption that this does not modify the results significantly (see, e.g., \cite{Acharya:2006ne, Douglas:2010rt, Blaback:2010sj, Saracco:2012wc, McOrist:2012yc, Gautason:2015tig} for discussions of the smearing issue).} The aim is to review the construction in an explicit string-theory example but also to discuss some problems in type IIA which we believe are not present in the type IIB models we will study in Sects.~\ref{sec:fluxwiniib} and \ref{sec:EFT_LF_AXION}. We will use the conventions of \cite{Palti:2008mg} and focus on the specific example of the mirror-dual of $\mathbb{CP}^4_{(1,1,2,2,6)}[12]$, which was analysed in \cite{Palti:2015xra} and illustrates the relevant points. The effective four-dimensional ${\cal N}=1$ supergravity is specified by the following Kahler  potential and superpotential:
\begin{align}
\mathcal{K} &= - 2 \mathrm{log} \left[\sqrt{s u_1} \left( u_2 - \frac23 u_1 \right) \right] + \mathcal{K}_T\left( T^A -\bar T^A\right) \,,  \\
\mathcal{W} &= - n_0 S - n_1 U_1 - n_2 U_2 + \mathcal{W}_T\left(T^A \right) + A_0\left(T^A\right) \mathrm{e}^{iS} + A_1\left(T^A\right) \mathrm{e}^{iU_1} + A_2\left(T^A\right) \mathrm{e}^{iU_2} \,. \label{kwiiacye}
\end{align}
Here, we introduced the complex-structure moduli $S = \sigma + is$, $U_1 =  a_1 + iu_1$, $U_2 = a_2 + iu_2$. The Kahler moduli $T^A$ have the Kahler potential
\begin{equation}
\mathcal{K}_T\left( T^A -\bar T^A \right)=-\log\left (\dfrac{i}{6}k_{ABC}(T^{A}-\bar T^{A})(T^{B}-\bar T^{B})(T^{C}-\bar T^{C})\right )
\end{equation}
and superpotential
\begin{align}
\cW_{T}(T^A)=\dfrac{f_{0}}{6}k_{ABC}T^{A}T^{B}T^{C}+\dfrac{1}{2}k_{ABC}\tilde{f}^{A}T^{B}T^{C}+f_{A}T^{A}+\tilde{f}_{0}\,,
\end{align}
where $f_{0}$, $\tilde{f}^{A}$, $f_{A}$ and $\tilde{f}_{0}$ are quantised RR fluxes, see, e.g., \cite{Grimm:2004ua}. We will ignore the stabilisation of the Kahler sector for the moment and come back to this point further below.

The terms linear in the complex-structure moduli in \eqref{kwiiacye} are induced by NSNS fluxes, while the exponential terms are instanton-induced. Although we have written instanton contributions for the moduli $S$, $U_1$ and $U_2$ in (\ref{kwiiacye}), the question of whether those contributions are present is a difficult one, cf.~\cite{Blumenhagen:2009qh} for a review. We will proceed with the assumption that they all contribute but will keep in mind that this may not be the case for specific examples.

Perturbatively, all saxions and one combination of the 3 axions $\sigma$, $a_1$ and $a_2$ are fixed.\footnote{The stabilisation scheme in this example is therefore slightly different from the rest of this paper where we always stabilise all but one axion combination by fluxes.} Non-perturbatively, all the axions gain a mass, but these masses can be exponentially different according to how the saxions $s$, $u_1$ and $u_2$ are fixed. In the vacuum, we have that \cite{Palti:2015xra}
\begin{equation}
\frac{s}{u_1} =\frac{3 n_1+2 n_2}{3 n_0} \,,\qquad  \frac{s}{u_2} = \frac{n_2\left( 3 n_1+2 n_2\right)}{6 n_0 \left( n_1+n_2\right)} \,,\qquad \frac{u_1}{u_2} = \frac{n_2}{2\left(n_1+n_2\right)} \,. \label{IIA-moduli-ratios}
\end{equation}
We now consider a setting where $n_0 \gtrsim n_1,n_2$. This implies that $s \lesssim u_1,u_2$. Of the three axions we therefore have two heavy combinations: $\sigma$ and $n_0 \sigma + n_1 a_1 + n_2 a_2$. The remaining (exponentially) light axion is orthogonal to both of these combinations. Since the metric on the field space factorises between $\sigma$ and the $a_i$, this light direction is purely in the $a_1$ and $a_2$ space and orthogonal to $n_1 a_1 + n_2 a_2$.

We therefore arrive at a situation similar to the two-axion toy model in the previous section. The periodicities of the two instantons associated to $u_1$ and $u_2$ can be written as
\begin{equation}
\Pi_1 = \frac{f}{p_2} \,,\qquad \Pi_2 =  \frac{f}{p_1} \,,\qquad f = \sqrt{\frac{3}{8}}\frac{p_2}{u_1} = \sqrt{\frac{3}{2}}\frac{p_1 + p_2}{u_2}\,.
\end{equation}
Here, as earlier, the $p_i$ are the co-prime factors in the $n_i$. We can now utilise this example to illustrate both aligned and misaligned winding. We see that aligned winding is realised for the choice
\begin{equation}
p_1 \sim {\cal O}\left(1\right) \,,\;\; p_2 \sim {\cal O}\left(N\right) \; \implies\; \Pi_1 \sim \frac{1}{u_1} \,,\;\; \Pi_2 \sim f \sim \frac{N}{u_1} \label{f1}
\end{equation}
for $N \gg 1$, while, for misaligned winding, we have
\begin{equation}
p_1 \sim {\cal O}\left(N\right) \,,\;\; p_2 \sim {\cal O}\left(N\right) \; \implies\; \Pi_1 \sim \Pi_2 \sim \frac{1}{u_1} \,,\;\;  f \sim \frac{N}{u_1} \,. \label{f2}
\end{equation}
Note that, due to the requirement of tadpole cancellation, $N$ is bounded by the number of O6-planes and can therefore not be made arbitrarily large.

It appears from the above discussion that, assuming the $u_i$ are not too large, we can use winding to construct super-Planckian field ranges in this model. However, there are in fact several problems:
\begin{itemize}
\item The example of aligned winding may at first appear to have parametrically large monotonic regions in the potential, associated to $\Pi_2$. However, it was observed in \cite{Palti:2015xra} that \eqref{IIA-moduli-ratios} implies $u_2 \approx 2 u_1$ for large $N$ and therefore the dominant instanton is the one associated to $\Pi_1$, which is not enhanced. This makes the above setup unsuitable for inflation (but does a priori not exclude that the model is still an example with parametrically large $f$).
\item A second, previously unnoticed problem is the backreaction of the Kahler moduli. In particular, in a controlled regime, they backreact on the vevs of the $u_i$ in such a way that the parametric enhancement of $f$ is cancelled.
This can be seen by noting that the energy densities in the 10d action have to be small in string units $2\pi\sqrt{\alpha^\prime}=1$, i.e.,
\begin{equation}
|H_3|^2 \lesssim 1\,, \quad \mathrm{e}^{2\phi}|F_p|^2 \lesssim 1\,,
\end{equation}
where $\mathrm{e}^\phi=g_s$ is the 10d dilaton and the contractions are with the string-frame metric. This ensures that higher-derivative corrections to the effective action are suppressed.\footnote{The dilaton factor in the second inequality is due to the usual definition of the RR fields with an extra power of $g_s$ (see, e.g., \cite{DeWolfe:2005uu}).} Consider now in particular the bound on the Romans mass, $\mathrm{e}^{2\phi}F_0^2 \lesssim 1$. In terms of the 4d moduli, this becomes
\begin{equation}
\frac{\mathcal{V}}{\sqrt{su_1}}\left(u_2-\frac{2}{3}u_1\right)^{-1} f_0^2 \lesssim 1\,, \label{pert2}
\end{equation}
where $\mathcal{V}$ is the string-frame volume and $f_{0}=F_{0}$ in string units.
This can be shown to follow from the usual definition of the complex-structure and Kahler moduli in type IIA \cite{Grimm:2004ua, DeWolfe:2005uu}.
Solving the $F$-term constraints for $s$ and $u_i$, we furthermore find, at leading order in $N$,
\begin{equation}
s \sim \frac{f_0 \mathcal{V}}{n_{0}}\kom u_i \sim \frac{f_0 \mathcal{V}}{N}\,, \label{sol}
\end{equation}
where we used that $\cW\sim \text{Im}\W_T \sim f_0 \mathcal{V}$ \cite{DeWolfe:2005uu}. Using \eqref{sol} in \eqref{pert2}, we arrive at the condition
\begin{equation}\label{eq:BVA} 
 \mathcal{V} \gtrsim N^2\, . 
\end{equation}
Replacing $\cV$ in \eqref{sol} by \eqref{eq:BVA}, we find
\begin{equation}
u_i \gtrsim f_0N\,.
\end{equation}
According to \eqref{f1}, \eqref{f2}, the effective axion decay constant for (mis-)aligned winding trajectories is $f \sim \frac{N}{u_1}$. Since $f_{0}$ is an integer, it follows that 
\begin{equation}
f \lesssim 1\,, \label{fff}
\end{equation}
and therefore the field range is necessarily small in the controlled regime.
\item We also stress that, in type IIA models, our light-axion EFT always lives in deep AdS space. This may sound surprising since, in the limit of large $F_4$ flux, the AdS curvature is small compared to the KK scale \cite{DeWolfe:2005uu}. We therefore have a genuinely four-dimensional low-energy EFT for all moduli in an approximate flat regime. However, one can also check that the AdS scale is always of the order of the saxion masses.\footnote{Depending on the compactification, some of the saxions and/or axions can be tachyonic, with masses above the Breitenlohner-Freedman bound \cite{DeWolfe:2005uu}.} To be able to integrate out the saxions, one needs to go to a lower scale into the deep AdS regime. It is therefore not possible to take a limit where a large-$f$ axion lives in approximate Minkowski space.
\end{itemize}

It is not clear to us whether there could be other type IIA models in which some or all of these issues are ameliorated. Also note that we assumed above that $n_0 \gtrsim n_1,n_2$ in order to arrive at an EFT for a single light axion, cf.~the discussion below \eqref{IIA-moduli-ratios}. One can show that, without this assumption, the bound \eqref{fff} is slightly relaxed to $f \lesssim \sqrt{N}/\sqrt{n_0}$. This is still smaller than the naive scaling $f\sim N$ but may at least allow a moderate enhancement $f \sim \sqrt{N}$ for small enough $n_0$. In order to verify this, one would have to study in more detail the EFT for the 2 axions orthogonal to the heavy combination $n_0\sigma+n_1a_1+n_2a_2$.

Notice further that the axions $a_{1}$, $a_{2}$ arising from the RR $3$-form do not suffer from loop corrections in type IIA. We emphasise this point because, as we will later see, loop corrections to the axion potential are a limiting factor in related type IIB models.

Nevertheless, we consider it more promising to study aligned and misaligned winding in type IIB string theory instead. As usual in type IIB, the stabilisation of the complex-structure moduli is approximately independent of the Kahler moduli due to a no-scale structure at tree-level, with only small corrections at sufficiently large volumes. The backreaction problem described above is therefore not expected to occur in such models. Indeed, we will discuss several candidate constructions in Sect.~\ref{sec:fluxwiniib} which plausibly realise large field ranges using the winding idea. We will also see in Sect.\ \ref{sec:EFT_LF_AXION} that, for sufficiently small $g_s$, the AdS scale is small compared to the moduli mass scale such that we have an approximate Minkowski situation.

\section{(Mis-)aligned winding in type IIB string theory}
\label{sec:fluxwiniib}

We now turn to type IIB compactifications, focusing again on complex-structure moduli and winding trajectories as discussed in \cite{Hebecker:2015rya}. The simplest setting is that of an effective-axion trajectory which is aligned with one axis in a two-axion plane. We find obstructions to realising large $f$ in this basic setting. Next, we suggest and analyse three loopholes: The first is based on misaligned winding as defined in Sect.\ \ref{sec:amis}. The second uses a finely tuned superpotential. The third relies on mixing between three axions. Finally, we attempt to make generic statements in a situation with any number of axions and a superpotential which is independent of one linear combination of these fields.

\subsection{A no-go theorem for aligned winding with two axions}
\label{sec:No_Go_Two_Axions_IIB}

As described in \cite{Hebecker:2015rya}, winding can be achieved with a flux-induced superpotential \cite{Giddings:2001yu} of the form \eqref{eq:W1},
\begin{equation}\label{eq:W1.1} 
\cW=w(Z)+f(Z)(n_{1}\, U_{1}+n_{2}\, U_{2})\, .
\end{equation}
Here, $U_{1}=a_{1}+i u_{1}$ and $U_{2}=a_{2}+i u_{2}$ are two distinguished complex-structure moduli. The remaining complex-structure moduli and the axio-dilaton are collectively denoted by the variable or set of variables $Z$. Without loss of generality, we assume that the fluxes $n_{1}$, $n_{2}$ are co-prime. Kahler  moduli stabilisation will be ignored for the moment -- it is discussed in Sect.~\ref{sec:EFT_LF_AXION}.

It is essential that we assume the moduli $U_{1}$, $U_{2}$ to be stabilised at large complex structure (LCS), such that $u_{1},u_{2}\gg 1$. As a result, terms in the periods of the CY which involve factors $\mathrm{e}^{iU_{1}}$ and $\mathrm{e}^{iU_{2}}$ can be ignored. This justifies the ansatz (\ref{eq:W1.1}). We are agnostic about the moduli $Z$ -- they may or may not be at LCS. We will assume, however, that their values are such that the Kahler  potential is still well-approximated by its leading term in the LCS expansion. For example, we want to avoid situations where terms like $i(U_i-\overline{U}_i)(U_j-\overline{U}_j) (Z-\overline{Z})\lesssim 1$ because we tuned $Z$ to achieve $|Z-\overline{Z}|\ll1$.

With these assumptions, the complex-structure sector of the Kahler  potential 
\begin{equation}
\label{eq:K1} \mathcal{K} \supset - \log \Big[ \mathcal{A}(Z, \bar{Z} , U_{1}-\bar{U}_{1}, U_{2}-\bar{U}_{2}) \Big]
\end{equation}
is shift-symmetric in $a_{1}=\R(U_{1})$ and $a_{2}=\R(U_{2})$. These are our axion candidates. More explicitly, the function $\cA$ has the structure \cite{4,3}
\begin{equation}\label{eq:AForIIB} 
\mathcal{A}(Z,\bar{Z},U_{1}-\bar{U}_{1}, U_{2}-\bar{U}_{2})=\cA_{3}(Z-\bar{Z},U_{1}-\bar{U}_{1}, U_{2}-\bar{U}_{2}) + i c +g(Z,\bar{Z})\,,
\end{equation}
which is best explained in the language of the mirror-dual 3-fold. In this language, $Z$ and $U_i$ are 2-cycle-related Kahler  moduli and $g(Z,\bar{Z})$ encodes worldsheet instanton effects $\sim\exp(iZ)$. The perturbative (in the dual language) term, which dominates at LCS or large dual volume, is given by the cubic polynomial
\begin{equation}\label{eq:AForIIBLCS} 
\cA_{3}(Z-\bar{Z},U_{1}-\bar{U}_{1}, U_{2}-\bar{U}_{2})=\frac{i}{3!} \kappa_{ijk} (Y^{i} - \bar{Y}^{i})(Y^{j} - \bar{Y}^{j})(Y^{k} - \bar{Y}^{k})\,.
\end{equation}
Here, $Y$ denotes both $Z$- and $U$-moduli. The $\kappa_{ijk}$ are dual intersection numbers and $c =-i\zeta(3) \chi(X_{3})/(4 \pi^3)$ with $\chi(X_{3})$ the Euler characteristic of the Calabi-Yau threefold $X_{3}$.

We now make a choice which is important for the following discussion: In the dual language, $Y^i$ are components of the Kahler  form in a certain basis. We choose this basis to be a basis of the Kahler  cone. As a result, the triple intersection numbers are non-negative integers, $\kappa_{ijk}\geq 0$, see, e.g., \cite{10,36}.

The key point for the winding idea is that the superpotential \eqref{eq:W1.1} is independent of a certain linear combination of $U_1$ and $U_2$. Hence, the $F$-term conditions 
\begin{equation}\label{eq:2Mod:FTerms}
D_{U_{1}}\W=n_{1}\, f(Z)+\cK_{U_{1}}\W=0\kom D_{U_{2}}\W=n_{2}\, f(Z)+\cK_{U_{2}}\W=0
\end{equation}
leave a flat direction on the $a_1$-$a_2$ field space. For $\W\neq 0$, these conditions can be rewritten as
\begin{align}
\label{eq:FTerm1} \dfrac{\cK_{U_{1}}}{\cK_{U_{2}}}&=\dfrac{n_{1}}{n_{2}}\,,\\
\label{eq:FTerm2} f(Z)+\dfrac{\cK_{U_{1}}}{n_{1}}\W&=0\, .
\end{align}
The first equation corresponds to fixing the relative volume of two (dual) $4$-cycles. Their overall volume is fixed by \eqref{eq:FTerm2}. This second equation also stabilises one linear combination of $a_1$ and $a_2$.

The plan is now to investigate properties of the field range $f$ of the surviving effective axion in the $a_1$-$a_2$-plane. This proceeds by analogy to the derivation of \eqref{fmixp}. First, recall the relevant kinetic terms,
\begin{align}
\cL&\supset \cK_{U_{1}\bar{U}_{1}}|\p U_{1} |^{2}+\cK_{U_{1}\bar{U}_{2}}(\p U_{1})(\p \bar U_{2})+h.c. +\cK_{U_{2}\bar{U}_{2}}|\p U_{2}|^{2}\; .\label{kuu}
\end{align}
Next, introduce the effective axion $\chi$ through $(a_{1},a_{2})\equiv(-n_{2},n_{1})\, \chi$, cf.~\eqref{eq:Def:Chi}. The decay constant of $\chi$ then reads
\begin{equation}\label{eq:FSquare1}
f^{2}=n_{2}^{2}\cK_{U_{1}\bar{U}_{1}}-2n_{1}n_{2}\cK_{U_{1}\bar{U}_{2}}  +n_{1}^{2}\cK_{U_{2}\bar{U}_{2}}\, .
\end{equation}

To analyse this result, it will be convenient to think in terms of derivatives with respect to the real variables $u_1$ and $u_2$. For example, we have $4\cK_{U_{1}\bar{U}_{2}}=\p_{u_{1}}\p_{u_{2}}\cK$ and $4\cK_{U_{2}}\cK_{\bar{U}_{2}}=(\p_{u_{2}}\cK)^{2}$. To simplify notation, we will furthermore write $2\cK_{2}\equiv \p_{u_{2}}\cK$, $4\cK_{12}\equiv \p_{u_{1}}\p_{u_{2}}\cK$ etc.\footnote{
Beware 
of the factor of $i$ appearing in the relation between, e.g., ${\cal K}_1$ and ${\cal K}_{U_1}$.
}
With this, we have for example
\begin{align}
\cK_{11}&=\cK_{1}\cK_{1}-\dfrac{\cA_{11}}{\cA}=\dfrac{n_{1}^{2}}{n_{2}^{2}}\cK_{2}\cK_{2}-\dfrac{\cA_{11}}{\cA}\,,
\end{align}
where we used \eqref{eq:FTerm1} in the second equality. Similar expressions can be given for $\cK_{12}$ and $\cK_{22}$. As a result, \eqref{eq:FSquare1} simplifies to
\begin{equation}\label{eq:FSquare_Final2Mod} 
f^{2}=-n_{2}^{2}\dfrac{\cA_{11}}{\cA}+2n_{1}n_{2}\dfrac{\cA_{12}}{\cA}-n_{1}^{2}\dfrac{\cA_{22}}{\cA}\, .
\end{equation}
Note that, while not apparent in this form, $f^2$ continues to be positive definite. In other words, in any consistent model, the various quantities on the right-hand side of \eqref{eq:FSquare_Final2Mod} will always take values which ensure positivity.

We would like to understand the properties of \eqref{eq:FSquare_Final2Mod} in case of aligned or misaligned trajectories. Let us first consider the case of a small ratio in \eqref{eq:FTerm1}. We choose fluxes $(n_{1},n_{2})\raw (1,N)$ with $N\gg 1$ corresponding to the setup investigated in \cite{Hebecker:2015rya}. Note that $N$ has to be positive because of \eqref{eq:FTerm1} and $\mathcal{K}_{U_1}/\mathcal{K}_{U_2}>0$. In this case, since only $a_{1}+N\, a_{2}$ is stabilised, we obtain a winding trajectory closely aligned with $a_{1}$, hence the name \textit{aligned} trajectory. The range of the $U_{1}$-axion is $N$-fold extended in this way. By~\eqref{eq:AForIIBLCS}, the function ${\cal A}$ is just a cubic polynomial in the imaginary parts of the $U_{i}$ with positive coefficients. Hence, its  second derivatives are non-negative: $\cA_{11},\cA_{12},\cA_{22}\geq 0$. Since $\cA>0$, the first and last term in \eqref{eq:FSquare_Final2Mod} contribute negatively. Thus, we estimate
\begin{equation}\label{eq:FSquare_2Mod_Est}
f^{2}\leq 2N \dfrac{\cA_{12}}{\cA}\, .
\end{equation}
We claim that this bound implies $f^{2} \ll 1$. 

To demonstrate this, rewrite the inequality as
\begin{align}\label{eq:FSquare_2Mod_Est_Re} 
f^{2}&\leq 2N\dfrac{u_{2}\cA_{2}}{\cA}\dfrac{\cA_{12}}{u_{2}\cA_{2}}=2N \dfrac{u_{2}\cA_{2}}{\cA} \dfrac{\cA_{12}}{u_{2}N\cA_{1}}\,,
\end{align}
where we used \eqref{eq:FTerm1} in the second step. Indeed, let us define the ratios
\begin{align}
\label{eq:E_fcts} E_{2}&\coloneqq\dfrac{\cA}{u_{2}\cA_{2}}\,,\qquad E_{12}\coloneqq \dfrac{\cA_{1}}{u_{2}\cA_{12}}\,.
\end{align}
In terms of $E_{2}$ and $E_{12}$, Eq.~\eqref{eq:FSquare_2Mod_Est_Re} becomes
\begin{equation}\label{eq:FSquare_2Mod_Est_Fin} 
f^{2}\leq \dfrac{2}{ u_{2}^{2}}\, \dfrac{1}{E_{2}E_{12}}\, .
\end{equation}
It is clear that, whenever $\mathcal{A}$ is dominated by the perturbative term \eqref{eq:AForIIBLCS}, we have $u_{2}{\cal A}_2 \leq  \cO(1){\cal A}$. The reason is simple: If all terms in ${\cal A}$ involve $u_2$, then $u_2{\cal A}_2\sim {\cal A}$ ignoring ${\cal O}(1)$ factors. But ${\cal A}$ may involve terms without $u_2$. These are annihilated when taking the derivative, thus making $u_2{\cal A}_2$ generically smaller. Therefore, the ratio $E_{2}$ cannot become parametrically small: $E_{2}\gtrsim 1$. This also holds for $E_{12}$ so that $E_{12}\gtrsim 1$. Since $u_{2}\gg 1$ in the LCS limit, we deduce that $f^{2} \ll 1$. We formulate our observations in terms of a no-go theorem:

\vspace{0.1cm}

\textbf{No-go theorem for aligned winding trajectories with two moduli:} \textit{Consider a IIB flux compactification on a Calabi-Yau orientifold. Let two complex-structure moduli $U_{1}$, $U_{2}$ be at LCS with all others such that the perturbative terms in the Kahler  potential dominate. If the superpotential $\cW$ only depends on $U_{1}$, $U_{2}$ through the linear combination}\footnote{Note that the no-go theorem does not rely on the specific form of the superpotential in Eq.~\eqref{eq:W1.1} but holds more generally for any $\W=\W(U_{1}+N U_{2},Z)$.} \textit{$U_{1}+N\, U_{2}$ with $N\gg 1$, then the field range of the remaining flat axionic direction cannot become parametrically large.\footnote{This no-go result does not apply to the toroidal examples of \cite{Hebecker:2017lxm}, which studied large field ranges in the regime $u_{i}\lesssim 1$. Due to the absence of instantons on the torus, this does not imply a loss of control.}}

\vspace{0.1cm}

\begin{figure}
\centering
 \includegraphics[width=0.5\textwidth]{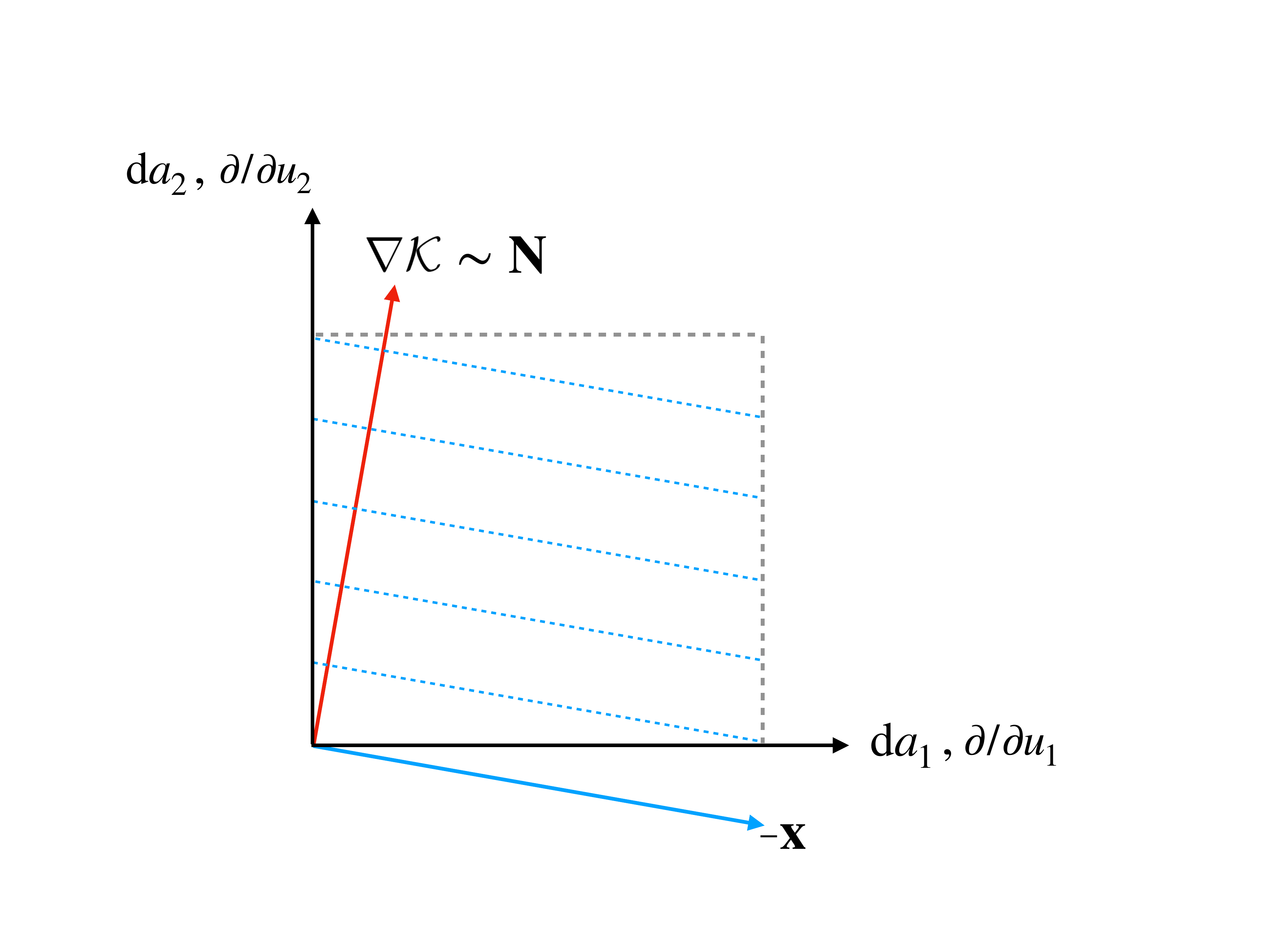}
\caption{Light axion direction (aligned with $a_1$ axis) vs. flux vector $\mathbf{N}$ and Kahler  potential gradient $\nabla \cK$ (aligned with orthogonal axis).}
\label{fig:2}
\end{figure}

To understand this result intuitively, rewrite the $F$-term conditions \eqref{eq:2Mod:FTerms} as
\begin{equation}
f(Z)\,\mathbf{N}-i\nabla\cK\,W=0 \label{fterm-vec}\,,
\end{equation}
with
\begin{equation} \label{def-nablak}
\mathbf{N}=\left (\begin{array}{c}
n_{1} \\ 
n_{2}
\end{array} \right )\kom \nabla\cK=\left (\begin{array}{c}
\cK_{1} \\ 
\cK_{2}
\end{array} \right )\,.
\end{equation}
Furthermore, recall that we parametrise the light axion direction by $\chi$ as in Sect.\ \ref{sec:amis} (see comment after (\ref{kuu})) and define a vector pointing into this direction,
\begin{equation}
\mathbf{x} = \frac{\text{d} \R(\mathbf{U})}{\text{d} \chi}, \qquad \mathbf{U} = \left (\begin{array}{c}
U_{1} \\ 
U_{2}
\end{array} \right )\,.
\end{equation}
Our intuitive argument is based on the arrangement of the three vectors $\mathbf{N}$, $\nabla\cK$ and $\mathbf{x}$:
On the one hand, $\nabla\cK$ lives in the tangent space of the saxionic field space, with basis $(\partial/\partial u_1, \partial/\partial u_2)$. The vector $\mathbf{N}$ is parallel to $\nabla\cK$ due to \eqref{fterm-vec} and can be plotted in the same space, cf. Fig.~\ref{fig:2}. On the other hand, $\mathbf{x}$ lives in the cotangent space of the axionic field space, with basis $(\mathrm{d}a_1,\mathrm{d}a_2)$. It will be convenient to identify these two vector spaces using the above bases. In other words, we draw all vectors in a single space, see again Fig.~\ref{fig:2}. Orthogonality between two vectors from these two spaces becomes Euclidean orthogonality. 

Now, the superpotential \eqref{eq:W1.1} forces the flat axionic direction to  satisfy $n_{1}\mathrm{d}a_{1}+n_{2}\mathrm{d}a_{2}=0$. In other words, the flat direction is orthogonal to the flux vector $\mathbf{N}$:
\begin{equation}
\mathbf{N}\cdot \mathbf{x}=0\,. \label{orth}
\end{equation}
Furthermore, $\nabla\cK \sim \mathbf{N}$, such that the direction of $\nabla \cK$ is fully determined by the light axion direction. In particular, aligning the light axion with the $a_1$-axis implies that $\nabla\cK$ is aligned with the $a_2$-axis, see Fig.~\ref{fig:2}. Because of $\nabla\cK=(\mathcal{K}_1, \mathcal{K}_2) \sim (\mathcal{A}_1,\mathcal{A}_2)$, this implies a large hierarchy $|\cA_{1}|\ll|\cA_{2}|$, which with \eqref{eq:AForIIBLCS} translates into a hierarchy between (combinations of) saxion vevs. This hierarchy makes $f$ small. 

To summarise, as we align $\chi$ with one of the fundamental axions, we are constrained to a special region in moduli space with a large hierarchy between the components of $\nabla\cK$. In that region, also the second-derivative matrix of $\cK$ is non-generic and, as shown by our no-go theorem, it counteracts the naive field-range extension due to $N\gg 1$. In the following, we will discuss possibilities to extend the field range by fluxes without being forced in such a special corner of moduli space (i.e., without the imposition of large hierarchies on the components of $\nabla\cK$). In this context, the geometric point of view introduced here will be very useful.

\subsection{Misaligned winding}
\label{sec:Mis-alignment_IIB} 

\begin{figure}
\centering
 \includegraphics[width=0.5\textwidth]{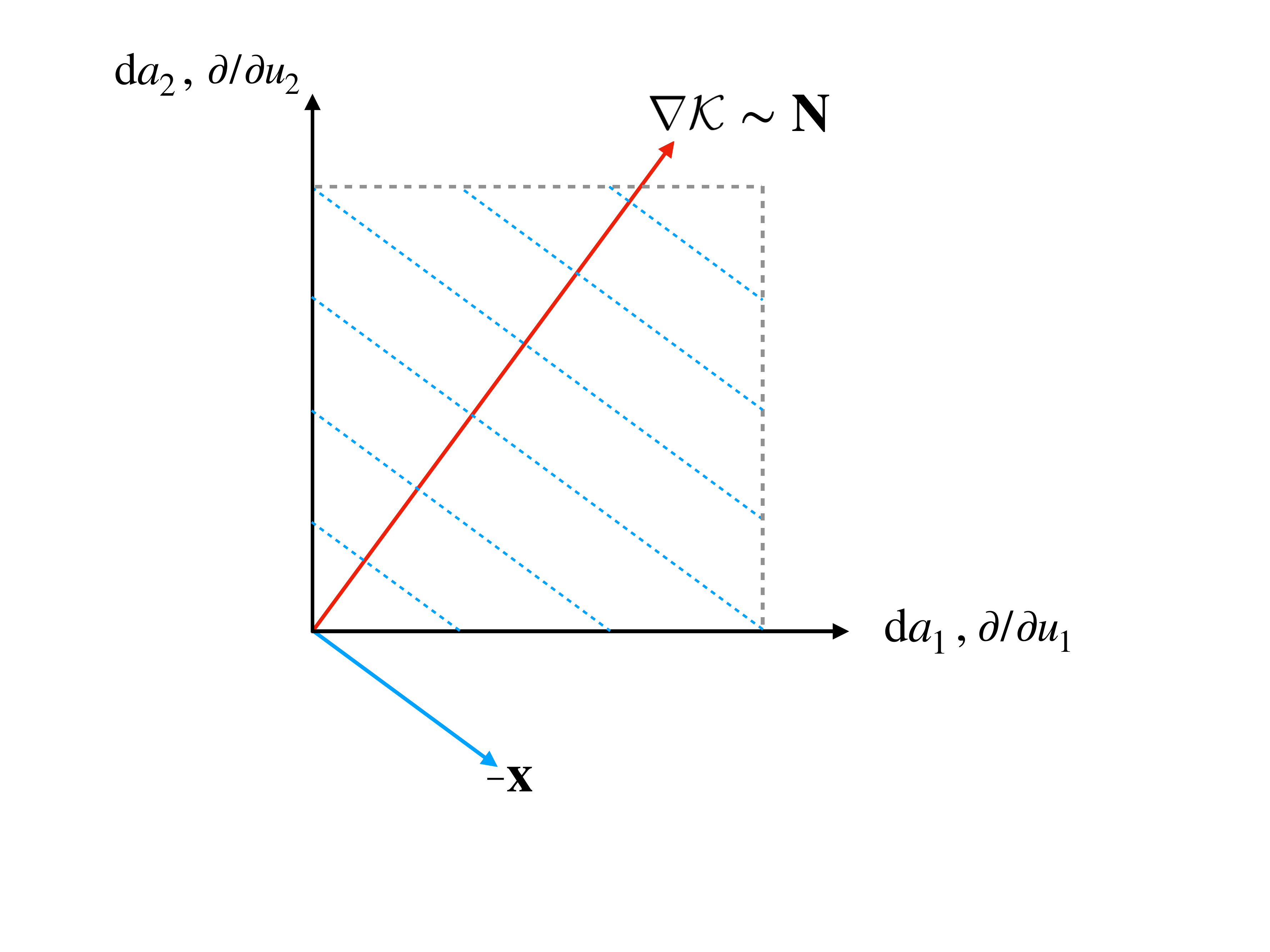}
\caption{Misaligned trajectory, labelled by $\mathbf{x}$, in two-moduli space. In this case, the vector $\nabla\cK$ is not aligned with any of the coordinate axes.}\label{fig:4}
\end{figure}

The winding trajectory discussed in the previous subsection was closely aligned with the $a_{1}$-axis. As discussed in Sect.\ \ref{sec:amis}, we can also think of different kinds of alignment, for example, with the diagonal direction $\mathrm{d}a_{1}-\mathrm{d}a_{2}$ (cf. Fig.~\ref{fig:4}). In this sense, the flat direction is {\it misaligned} with the original axions $a_{1}$, $a_{2}$. This can be achieved by the flux choice $(n_{1},n_{2})=(N,N+1)$ with $N\gg 1$. The  $F$-term constraint \eqref{eq:FTerm1} then becomes
\begin{equation}
\dfrac{\cK_{U_{1}}}{\cK_{U_{2}}}=\dfrac{\cA_{U_{1}}}{\cA_{U_{2}}}=\dfrac{N}{N+1}\sim\cO(1)\,.
\end{equation}
Contrary to the alignment scenario, this does not impose a large hierarchy between moduli vevs. Repeating the steps after \eqref{eq:FSquare_Final2Mod} for the flux choice $(n_{1},n_{2})=(N,N+1)$, we furthermore find that the bound on the axion decay constant is relaxed,
\begin{equation}
f^{2}\leq 2\, N(N+1) \dfrac{\cA_{12}}{\cA} = \frac{2}{u_2^2}\frac{N^2}{E_2E_{12}}\,.
\end{equation}
The key point here is the enhancement factor $N^2$ relative to the aligned case of~\eqref{eq:FSquare_2Mod_Est_Fin}. This evades the no-go theorem.

As in the aligned case, a roughly $N$-fold enhancement of the axion periodicity arises from the winding around the torus of the original field space. However, in contrast to the aligned case, this enhancement is not counteracted by a hierarchy between moduli vevs and, hence, between $\cA_i$ and $\cA_{ij}$ factors. 

The corresponding geometry is illustrated in Fig.~\ref{fig:4}: As $\mathbf{x}$ is aligned with one diagonal in the axion field space, $\nabla \cK \sim \mathbf{N}$ must be aligned with the opposite diagonal in the saxionic tangent space (cf.~\eqref{orth}). Thus, we immediately see that there is no hierarchy between the components of $\nabla \cK$. As explained above, we believe that this is the underlying reason allowing for long trajectories.

\subsection{Fine-tuned superpotential}

Our no-go theorem of Sect.~\ref{sec:No_Go_Two_Axions_IIB} excludes aligned trans-Planckian trajectories in two-axion models. A key assumption in its derivation was the use of a purely perturbative (in the mirror-dual language) Kahler  and superpotential with respect to the moduli $U_1$ and $U_2$. Thus, a natural way out might be to generalise our superpotential~\eqref{eq:W1.1} by including instanton terms:
\begin{equation}\label{eq:W2} 
\cW=w(Z)+f(Z)(n_{1}\, U_{1}+n_{2}\, U_{2})+A_1(Z)\mathrm{e}^{i U_{1}}+A_2(Z)\mathrm{e}^{i U_{2}}\,.
\end{equation}
The $F$-term constraints \eqref{eq:FTerm1}, \eqref{eq:FTerm2} become
\begin{align}
\dfrac{\cK_{U_{1}}}{\cK_{U_{2}}}&=\dfrac{n_{1}\, f(Z)+i A_1(Z)\mathrm{e}^{i U_{1}}}{n_{2}\, f(Z)+i A_2(Z)\mathrm{e}^{i U_{2}}}\,, \label{eq:RatioK1K2TowInf1}  \\
f(Z)&=-i \frac{A_1(Z)\mathrm{e}^{i U_{1}}}{n_1}-\dfrac{\cK_{U_{1}}}{n_{1}}\W\,. \label{eq:RatioK1K2TowInf2}
\end{align}
As in Sect.~\ref{sec:No_Go_Two_Axions_IIB}, we choose $(n_1,n_2) = (1,N)$ with $N\gg 1$ and assume $u_1,u_2 \gg 1$. The idea is now to tune the function $f(Z)$ to be very small, such that the exponentially suppressed instanton terms in~\eqref{eq:RatioK1K2TowInf1} can compete with it. By~\eqref{eq:RatioK1K2TowInf2}, $\W$ and hence $w(Z)$ will then also be small. Thus, in spite of the hierarchy between $n_1$ and $n_2$, we can hope to arrange $\cK_{U_1}/\cK_{U_2}\sim \cO(1)$. As argued above, a large hierarchy between the components of $\nabla \cK$ was the key issue underlying our no-go result. This suggests that large field ranges can be realised in models with a fine-tuned superpotential.

A possible objection to this construction is that the key superpotential term $f(Z)(U_1+NU_2)$, which is responsible for stabilising the 
axion combination $a_1+Na_2$, is tuned small. One might be concerned that the hierarchy between this stabilised field and the light axion $\chi$ will be lifted due to the tuning of $f(Z)$. Clearly, this would go against the spirit of the whole approach. However, we do not expect this to be a problem in general. Indeed, let us assume for notational simplicity that $Z$ stands for just one modulus and consider the corresponding $F$-term contribution to the scalar potential:
\be
|D_Z \W|^2= |\partial_Z f(Z)|^2 \,|U_1+NU_2|^2 + \partial_{Z} f(Z)(U_1+N U_2) \K^{Z\bar Z} \partial_{\bar Z} \bar w(\bar Z) + c.c. + \ldots\,,
\ee
where we dismissed exponentially small terms and terms that do not depend on $U_1+NU_2$. In order that the flux-induced mass generated for $a_1+Na_2$ does not become small, it is mandatory that the tuning for $|f|\ll 1$ does not imply $|\partial_Z f|\ll 1$.
While we cannot exclude obstructions due to the $F$-term constraints for the $Z$-moduli, generic flux choices do not imply $|\partial_Z f|\ll 1$:
For example, if $f(Z)\equiv Z$ becomes small because we are stabilised near the locus $Z=0$ in moduli space, $\partial_Zf$ continues to be unity.

Another potential worry is that it may not be possible to arrange both $u_1,u_2 \gg 1$ and $f(Z) \ll 1$ in a given compactification. In particular, one might be concerned that the $F$-term constraints for the $Z$-moduli restrict the allowed on-shell values of $w(Z)$ and/or $f(Z)$ such that \eqref{eq:RatioK1K2TowInf2} cannot be solved with $u_1,u_2 \gg 1$. Indeed, this turns out to be the case in simple models (e.g., on the torus). However, we see a priori no reason why this should be a general issue. In particular, in compactifications with several $Z$-moduli, we expect to have enough tuning freedom to realise the above conditions.

To make this point clearer, we want to review the fine-tuning cost of the presented construction. As before, we are dealing with a flux superpotential $\W(Z,U_1,U_2)$ with $Z=\{Z_1,\cdots,Z_n\}$. We have to make a flux choice ensuring the particular structure $\cW\supset f(Z)(U_1+NU_2)$. The remaining flux choice is used, as is standard in the type IIB landscape framework, to place ourselves at a particular locus in complex-structure moduli space. The well-known underlying idea is that, via the solution of the $F$-term equations, the flux discretuum is mapped to an (in general rather dense) discretuum of points in moduli space (see, e.g.,~\cite{Denef:2004ze}). In this discretuum, we have to choose a point with $u_1,\,u_2\gg 1$ and $|f(Z)|\ll 1$. Only the last tuning is special to the present subsection. The smallness of $\W$ and of $w(Z)$ follows from the $F$-term equation~\eqref{eq:RatioK1K2TowInf2} and the definition of $w$ in~\eqref{eq:W2} and requires no further tuning.

\subsection{Generalisation to three axions}\label{sec:ThreeAxions} 

So far, we have formulated a no-go theorem for aligned winding trajectories and discussed two loopholes in scenarios where two axions mix. A third way of evading our no-go theorem is to consider the mixing of more than two axions. Indeed, we find evidence that already the mixing of three axions is sufficient to allow for long trajectories. As we will see, this is nicely illustrated by our geometric picture developed earlier.

Consider the superpotential
\begin{equation}\label{eq:3Mod:W} 
\cW=w(Z)+f_{1}(Z)(n_{1}\, U_{1}+n_{2}\, U_{2}+n_{3}\, U_{3})+f_{2}(Z)(m_{1}\, U_{1}+m_{2}\, U_{2}+m_{3}\, U_{3})
\end{equation}
as a simple generalisation of \eqref{eq:W1.1}. The Kahler  potential is defined as in \eqref{eq:K1}, \eqref{eq:AForIIB} with the replacement $\mathcal{A}(Z, \bar{Z} , U_{1}-\bar{U}_{1}, U_{2}-\bar{U}_{2})\to \mathcal{A}(Z, \bar{Z} , U_{1}-\bar{U}_{1}, U_{2}-\bar{U}_{2}, U_{3}-\bar{U}_{3})$. Our superpotential now only depends on two linear combinations of the three moduli $U_{i}=a_{i}+i u_{i}$. The Kahler  potential involves just the imaginary parts $u_{i}$. Thus, our setup has one light axion, just as in the case with only two distinguished moduli $U_1$ and $U_2$.

The $F$-term conditions for the $U_{i}$ read
\begin{align}\label{eq:3Mod:FTerms}
D_{U_i}\cW&=n_{i} f_{1}+m_{i}f_{2}+\cK_{U_i}\cW=0\qquad\forall\, i=1,2,3\; .
\end{align}
For $\W\neq 0$, this fixes the ratios of Kahler  potential derivatives according to
\begin{align}\label{eq:3Mod:FTerms_Ra_Gen} 
\dfrac{\cK_{U_j}}{\cK_{U_i}} = \frac{\mathcal{A}_{U_j}}{\mathcal{A}_{U_i}}&=\dfrac{n_{j}f_{1}+m_{j}f_{2}}{n_{i}f_{1}+m_{i}f_{2}}\qquad\forall\, i,j=1,2,3\; .
\end{align}
We need the kinetic term of the light axion obtained after integrating out the heavy axionic combinations $n_{i}a_{i}$, $m_{i}a_{i}$ as well as the saxions $u_i$. As before, we parametrise the light axionic direction by a field $\chi$:
\begin{equation}\label{eq:Def:Chi3Mod} 
\left (\begin{array}{c}
a_{1} \\ 
a_{2} \\ 
a_{3}
\end{array} \right )=\left (\begin{array}{c}
x_{1} \\ 
x_{2} \\ 
x_{3}
\end{array} \right )\chi\qquad\mbox{or}\qquad \mathbf{a}=\mathbf{x}\,\chi\,.
\end{equation}
Here, $\mathbf{x}$ is the smallest integer vector orthogonal to the flux vectors $\mathbf{N}=(n_1,n_2,n_3)^T$ and $\mathbf{M}=(m_1,m_2,m_3)^T$. This ensures that the field $\chi$ is $2\pi$-periodic. Explicitly, we have 
\begin{equation}\label{eq:3Mod:Flux_Comb} 
x_{1}=\frac{n_{3}m_{2}-n_{2}m_{3}}{x_0}\kom x_{2}=\frac{n_{1}m_{3}-n_{3}m_{1}}{x_0}\kom x_{3}=\frac{n_{2}m_{1}-n_{1}m_{2}}{x_0}
\end{equation}
with
\begin{equation}
x_0 = \text{gcd}\left( n_{3}m_{2}-n_{2}m_{3}, n_{1}m_{3}-n_{3}m_{1}, n_{2}m_{1}-n_{1}m_{2} \right)\,.
\end{equation}
A calculation similar to that in Sect.\ \ref{sec:No_Go_Two_Axions_IIB} determines the axion decay constant
\begin{align}\label{eq:3Mod:FSquare_Fin} 
f^{2}&=\dfrac{1}{\cA}\biggl \{-2\cA_{31}  x_{1}x_{3}- 2\cA_{32}  x_{2}x_{3}-2 \cA_{12} x_{1}x_{2} -\sum_{k=1}^{3}\, \cA_{kk}x_{k}^{2}\biggl \}\, .
\end{align}
The periodicities with respect to each of the three fundamental instantons are
\begin{equation}\label{eq:3Mod:PeriodicitiesInstantons} 
\Pi_{1}=\dfrac{f}{x_{1}}\kom\Pi_{2}=\dfrac{f}{x_{2}}\kom\Pi_{3}=\dfrac{f}{x_{3}}\, .
\end{equation}
Again, we observe that the diagonal terms $\sim \cA_{kk}$ always contribute negatively to $f^{2}$. Therefore, we estimate
\begin{align}\label{eq:3Mod:FSquare_Est} 
f^{2}&\leq \dfrac{2}{\cA}\biggl \{ -\cA_{31}  x_{1}x_{3}- \cA_{32}  x_{2}x_{3}- \cA_{12} x_{1}x_{2} \biggl \}\, .
\end{align}
Of course, any physical configuration must lead to $f^{2}>0$. As in the two-axion case, this is not manifest in \eqref{eq:3Mod:FSquare_Fin}, \eqref{eq:3Mod:FSquare_Est} but is ensured by the consistency of the underlying model. 

The key observation is now that, contrary to the two-axion case discussed in Sect.\ \ref{sec:No_Go_Two_Axions_IIB}, it is not possible to derive a no-go theorem against large field trajectories using the ratios \eqref{eq:3Mod:FTerms_Ra_Gen}. In particular, the no-go argument of Sect.\ \ref{sec:No_Go_Two_Axions_IIB} involved using \eqref{eq:FTerm1} in \eqref{eq:FSquare_2Mod_Est} such that the flux dependence cancelled out in $f^2$ and a bound $f^2 < 1$ could be obtained. One can convince oneself that an analogous argument cannot be made in the three-axion case, i.e., trying to rewrite \eqref{eq:3Mod:FSquare_Est} using \eqref{eq:3Mod:FTerms_Ra_Gen} cannot lead to a bound due to the more complicated dependence on the fluxes $n_i$, $m_i$. We conclude that the aligned winding scenario with three axions is less restrictive than the two-axion version such that we may hope to realise large trajectories in models of this type.

\begin{figure}
\centering
 \includegraphics[width=0.6\textwidth]{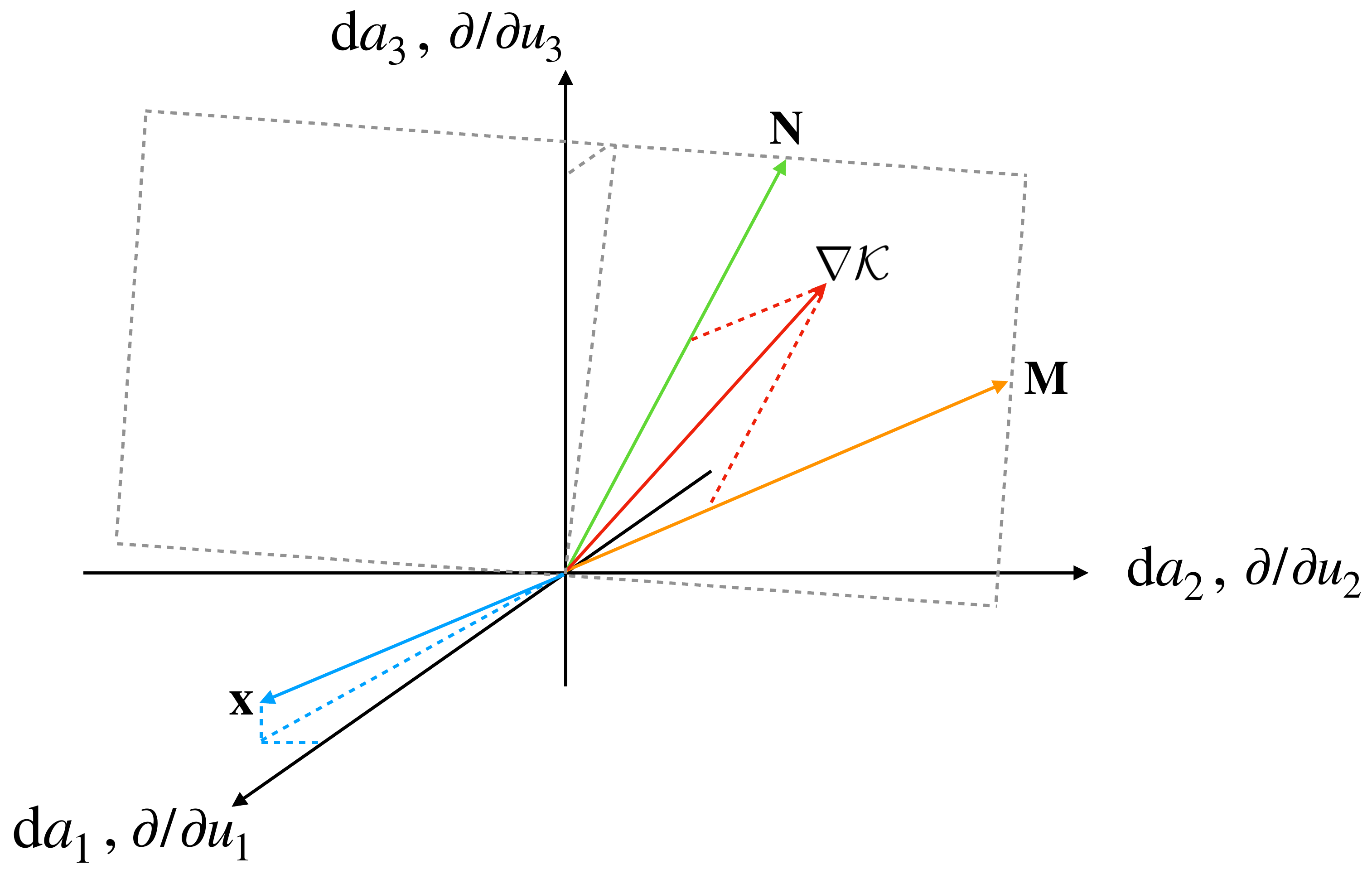}
\caption{The light axionic direction $\mathbf{x}$ is aligned with the $a_{1}$-axis. The hierarchy $|\cA_{1}|\ll|\cA_{2}|,|\cA_{3}|$ is apparent since the plane spanned by the flux vectors is nearly parallel to the $a_2$-$a_3$-plane.}\label{fig:1}
\end{figure}

Clearly, the failure of the old logic does not imply that things are actually better. To gain more confidence, a simple, intuitive understanding of the advantages of the three-axion case over the two-axion case is needed. Such an understanding can indeed be gained using the geometric interpretation of (mis-)aligned winding established earlier. We need to extend this picture to the present scenario. To do so, rewrite the $F$-term constraints \eqref{eq:3Mod:FTerms} as
\begin{equation}\label{eq:3Mod:Geo_K} 
\mathbf{N}\, f_{1}+\mathbf{M}\, f_{2}-i (\nabla\cK)\,\cW=0\,,
\end{equation}
where
\begin{equation}
\mathbf{N}=\left (\begin{array}{c}
n_{1} \\ 
n_{2} \\ 
n_{3}
\end{array} \right )\kom \mathbf{M}=\left (\begin{array}{c}
m_{1} \\ 
m_{2} \\ 
m_{3}
\end{array} \right )\kom \nabla \cK=\left (\begin{array}{c}
\cK_{1} \\ 
\cK_{2} \\ 
\cK_{3}
\end{array} \right ).
\end{equation}
We see that $\nabla\cK$ lies in the plane spanned by $\mathbf{N}$ and $\mathbf{M}$: 
\begin{equation}\label{eq:3Mod:Geo_KA} 
\nabla\cK\sim \mathbf{N}\, f_{1}+\mathbf{M}\, f_{2}\,.
\end{equation}
As stated above, the vector $\mathbf{x}$ satisfies
\begin{align}\label{eq:3Mod:Geo_Psi}
\mathbf{N}\cdot \mathbf{x}=0\kom \mathbf{M}\cdot \mathbf{x}=0\,,
\end{align}
which means that $\mathbf{x}$ is orthogonal to the plane spanned by $\mathbf{N}$ and $\mathbf{M}$.

First, consider the configuration in Fig.~\ref{fig:1}. Here, the light axionic direction is closely aligned with the $a_{1}$-axis. As a result, the plane of allowed values of $\nabla\cK$ is almost parallel to the $a_{2}$-$a_{3}$-plane. This induces a hierarchy of the form $|\cA_{1}|\ll|\cA_{2}|,|\cA_{3}|$. Such a hierarchy again translates into a hierarchy of the different moduli involved. Consequently, we expect an obstruction to large field ranges.
Indeed, it is straightforward to see this, analogously to the no-go argument of Sect.\ \ref{sec:No_Go_Two_Axions_IIB}: Since $\mathbf{x}$ is aligned with the $a_1$-axis, we have $x_1$ large while $x_2,x_3 \sim \mathcal{O}(1)$. From $\nabla\K \cdot \mathbf{x} =0$, it follows $\sum_i \mathcal{A}_i x_i =0$ and hence $x_1 = -(\mathcal{A}_2x_2+\mathcal{A}_3x_3)/\mathcal{A}_1$. Substituting this into \eqref{eq:3Mod:FSquare_Est} and using that the ratios $\mathcal{A}_{i1}/\mathcal{A}_{1}$, $\mathcal{A}_{i}/\mathcal{A}$ and $\mathcal{A}_{32}/\mathcal{A}$ are all small, we then find $f^2 \ll 1$ as previously.
Aligning the light axion direction with one of the $a_i$-axes therefore does not lead to long trajectories.\footnote{The same argument applies if one considers a mixing of more than three axions.}

\begin{figure}[t]
\centering
 \includegraphics[width=0.6\textwidth]{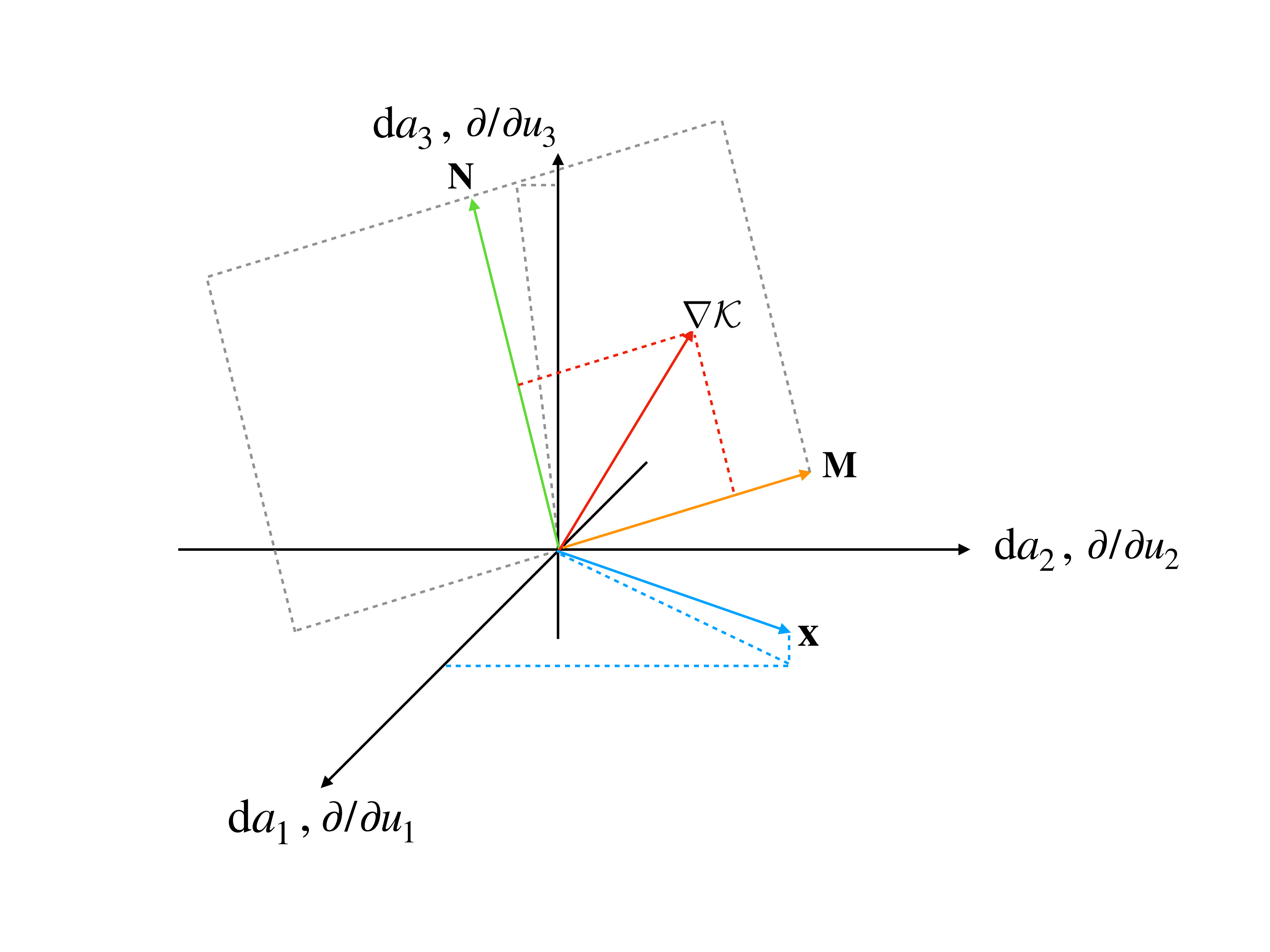}
\caption{Illustration of a flux choice for which $\mathbf{x}$ is nearly aligned with a generic direction of the $a_{1}$-$a_{2}$-plane. The plane of allowed values of $\nabla\cK$ is orthogonal to $\mathbf{x}$ and contains vectors without any hierarchy between their components.}\label{fig:3}
\end{figure}

However, and this is the crucial punchline of this subsection, much more promising geometries also exist. Indeed, consider the setup of Fig.~\ref{fig:3}, where $\mathbf{x}$ is nearly aligned with the $a_{1}$-$a_{2}$-plane, but not with one of its axes. The plane of allowed values of $\nabla\cK$ is still non-generic: It almost contains the $a_3$-axis. However, this plane now contains vectors $\nabla\cK$ which lie generically in the coordinate system -- they do not need to be aligned with any of the axes or planes. The freedom of choosing such a vector is in the coefficients $f_{1}$ and $f_{2}$ in Eq.~\eqref{eq:3Mod:W}. Thus, the ratios of the $\cA_{i}$ can in principle all be $\cO(1)$. We expect no obstructions to realising large $f$. 

Crucially, this last scenario falls in the category of aligned rather than misaligned winding. In particular, if the moduli can be stabilised such that $\mathrm{e}^{-u_1},\,\mathrm{e}^{-u_2}\ll \mathrm{e}^{-u_3}$ then, from the perspective of the axionic field $\chi$, the instanton with the long period $\Pi_3$ dominates those with the short periods $\Pi_1$ and $\Pi_2$. As a result, a violation of the strong form of the WGC appears possible. In order to use this for large-field inflation, additional constraints need to be satisfied, which will be discussed in Sect.~\ref{sec:Towards_Inflation}. Note also that achieving this hierarchy between the instantons does not imply a related hierarchy between the saxion values. Indeed, the exponentiation ensures that even an ${\cal O}(1)$ ratio between, say, $u_1$ and $u_3$ is sufficient to completely suppress the former instantons.

\subsection{General analysis at large complex structure}\label{sec:General_Argument} 

In the preceding subsections, we discussed several proposals to engineer long winding trajectories in spite of the no-go theorem of Sect.\ \ref{sec:No_Go_Two_Axions_IIB}. The aim of the current subsection is to test the winding scenario in a concrete setting where all $F$-term constraints can be solved explicitly. In particular, we will study Calabi-Yau compactifications in the limit of large complex structure, where the Kahler  potential takes a rather simple form. We will admit a completely general type IIB flux superpotential and allow mixing of an arbitrary number of axions. The goal is to determine the conditions under which aligned or misaligned trajectories can arise in this setting.

Concretely, we consider compactifications with $n=h^{2,1}$ complex-structure moduli $Y^i$. As before, we assume that the flux superpotential only depends on $n-1$ linear combinations of them,
\begin{equation}
\W = \W(n_{ai}Y^i, S)\,, \label{gen-superpotential}
\end{equation}
where $n_{ai}$ with $a=1,\ldots,n-1$ are integer flux numbers and $S$ is the axio-dilaton. The Kahler  potential is given by
\begin{equation}
\mathcal{K} = -\text{log} \mathcal{A} -\text{log} \left[ -i (S - \bar{S})\right] + \cK_T \,, \quad \mathcal{A}= \frac{i}{3!} \kappa_{ijk} (Y^{i} - \bar{Y}^{i})(Y^{j} - \bar{Y}^{j})(Y^{k} - \bar{Y}^{k})\,. \label{ka}
\end{equation}
Here, $\cK_T$ denotes the part depending on the Kahler  moduli, which need to be stabilised by quantum effects (see Sect.\ \ref{sec:EFT_LF_AXION}). It will be irrelevant for the current discussion, where we will focus on the tree-level stabilisation of the complex-structure moduli and the axio-dilaton.

In the large-complex-structure limit, $\cK$ is shift-symmetric and satisfies a no-scale condition (see, e.g., \cite{Grimm:2004ua}). This implies the useful relations
\begin{align}
\text{Im}(Y^i) &= \frac{i}{2}\cK^i\,, \quad
\cK_i \text{Im}(Y^i) = \frac{3i}{2}\,, \quad
\cK_{ij} \text{Im}(Y^i) = \frac{i}{2} \cK_j\,, \quad
\cK_{ijk} \text{Im}(Y^i) = i \cK_{jk}\,. \label{gsgtsg}
\end{align}
We also have
\begin{align}
\cK_{ij} &= \cK_{i} K_j - \frac{\mathcal{A}_{ij}}{\mathcal{A}}\,. \label{noscale1}
\end{align}
Here and in the remainder of this section, it will be convenient to use the standard notation where indices on $\K$ denote derivatives with respect to the complex fields, i.e., $\K_i \equiv \partial_{Y^i}\K$, $\K_{\bar \imath} \equiv \partial_{\bar Y^{i}}\K$ and analogously for $\W$, $\mathcal{A}$, etc.\footnote{Note that this differs from our notation in the previous subsections, where indices denoted derivatives with respect to real fields, cf.\ the comment below \eqref{eq:FSquare1}. These conventions differ by factors of $i$.} Note that, due to the shift symmetry, barred indices on $\K$, $\mathcal{A}$ can be replaced by unbarred ones using minus signs, e.g., $\K_{\bar\imath} = -\K_i$, $\K_{i \bar\jmath} = -\K_{ij}$, etc.

Let us parametrise the complex field direction along which $\W$ is constant by $X$, with $\text{Re}(X)=\chi$. This field direction is then given by $Y^i = x^i X$, where we define
\begin{equation}\label{eq:DefX} 
x^i=\frac{\text{d} Y^i}{\text{d} X} = \frac{\text{d} \text{Re}Y^i}{\text{d} \chi}
\end{equation}
as the smallest integer vector orthogonal to the vectors $n_{ai}$, in analogy to the previous sections. Since $\W$ only depends on $n-1$ combinations of the $Y^i$ and $\cK$ only depends on their imaginary parts, the $F$-term scalar potential generically leaves us with one light axion $\chi$.

Since $\W$ is independent of $X$ by assumption, the $F$-term conditions
\begin{align}
D_X \W &= \K_X \W = 0
\end{align}
imply, for $\W\neq 0$, that $\K_X=\mathcal{A}_X=0$ on-shell. The decay constant of the light axion $\chi$ is therefore
\begin{equation}
f^2 = \K_{X\bar X} = - \frac{\mathcal{A}_{X\bar X}}{\mathcal{A}} = -\frac{\mathcal{A}_{i\bar\jmath}}{\mathcal{A}} x^i x^{\bar\jmath}\,. \label{fieldrange1}
\end{equation}

Our goal is now to determine the conditions imposed on $x^i$ by the $F$-term constraints in order to assess under which circumstances large field ranges are possible in this setting.
To this end, we consider the general type IIB flux superpotential \cite{Gukov:1999ya, Giddings:2001yu}
\begin{equation}
\W = w(S) + f_i(S) Y^i + \kappa_{ijk}g^i(S) Y^j Y^k + h(S) \kappa_{ijk}Y^i Y^j Y^k + 3 h(S) c \label{w}
\end{equation}
with $c\sim i \chi(X_3)$ as in Sect.~\ref{sec:No_Go_Two_Axions_IIB} and
\begin{equation}
w(S)=w_0+w_1S\,, \quad f_i(S)=f_{0i}+f_{1i} S\,, \quad g^i(S)=g_0^i+g_1^iS\,, \quad h(S)=h_0+h_1 S\,.
\end{equation}
Here, $w_\alpha,f_{\alpha i},g_\alpha^i,h_\alpha$ with $\alpha=0,1$ are numbers given by sums involving integer flux numbers and classical intersections (see, e.g., (A.16) in \cite{Cicoli:2013cha}). For convenience, we will temporarily assume in the following that these numbers (and, accordingly, also the vector $x^i$) can take any real value and postpone a discussion of flux quantisation to the end of the section.

In order to bring \eqref{w} into the form \eqref{gen-superpotential}, we need to impose that $\W_X=\W_i x^i=0$ holds off-shell, which translates into a number of conditions on the fluxes. We first observe that we require
\begin{equation}
h(S) = 0\,. \label{hs}
\end{equation}
This follows because, for $h(S) \neq 0$, $\W_X=0$ would imply
\begin{equation}
\kappa_{ijk}x^k = 0 \quad \forall i,j\,. \label{wphi1}
\end{equation}
However, this would not be compatible with a non-vanishing axion decay constant
\begin{equation}
f^2 \propto \mathcal{A}_{X\bar X} \propto \kappa_{ijk}\text{Im}(Y^i)x^jx^k \neq 0 \label{aphiphi}
\end{equation}
such that we have to impose \eqref{hs} as claimed.
The requirement $\W_X=0$ also constrains those terms in \eqref{w} that are quadratic or linear in the $Y^i$. We thus find the conditions
\begin{equation}
\kappa_{ijk}g_\alpha^j x^k = 0 \quad \forall i\,, \qquad f_{\alpha i} x^i = 0\,. \label{wphi3}
\end{equation}
This is analogous to the orthogonality conditions we had in previous subsections (cf., e.g., \eqref{eq:3Mod:Geo_Psi}). These are $2n+2$ real homogeneous conditions for $n$ components of the real vector $x^i$.
Since we want exactly one light axion combination, the direction of $x^i$ should be completely fixed by the fluxes $f_{\alpha i},g_\alpha^i$. We therefore demand that $n-1$ conditions in \eqref{wphi3} are linearly independent. Thus, \eqref{wphi3} determines $x^i$ up to an overall scale.

We will now show how to solve the $F$-term constraints for the above setup. For convenience, we will set all axion vevs to zero, i.e., $\text{Re}(Y^i)=\text{Re}(S)=0$. This can always be done without loss of generality since the axions are shift-symmetric up to a change in the flux numbers. We can therefore absorb any axion vevs into the flux parameters in $\W$ (recall that we temporarily neglect flux quantisation). Furthermore, it will be crucial that we will solve the $F$-term constraints for the fluxes instead of the moduli. This has the advantage that all constraints can in the end be written as a rather simple equation system.

The $F$-term constraints $D_i \W = D_S\W =0$ yield
\begin{align}
0 &= f_{0i} - 2\kappa_{ijk} g_1^j \text{Im}(S) \text{Im}(Y^k) - 2 \frac{\text{Im}(\W)}{\mathcal{A}} \kappa_{ijk}\text{Im}(Y^j)\text{Im}(Y^k)\,, \\
0 &= w_1- \kappa_{ijk}g_1^i\text{Im}(Y^j)\text{Im}(Y^k) - \frac{\text{Im}(\W)}{2\text{Im}(S)}\,, \\
0 &= f_{1i} \text{Im}(S) + 2\kappa_{ijk} g_0^j \text{Im}(Y^k) + 2 \frac{\text{Re}(\W)}{\mathcal{A}} \kappa_{ijk}\text{Im}(Y^j)\text{Im}(Y^k)\,, \label{fterm-cs} \\
0 &= f_{1i} \text{Im}(Y^i) + \frac{\text{Re}(\W)}{2\text{Im}(S)}\,. \label{fterm-dil}
\end{align}
One can check that the terms in the first two equations are all proportional to one of the fluxes $w_1$, $f_{0i}$, $g_1^i$. For simplicity, we will solve them by setting
\begin{equation}
w_1=f_{0i}=g_1^i=0\,.
\end{equation}
The only non-trivial equations are then \eqref{fterm-cs}, \eqref{fterm-dil}. Analysing these equations will be sufficient to illustrate our main points, which can easily be generalised to solutions with non-zero $w_1$, $f_{0i}$, $g_1^i$.

In order to further simplify \eqref{fterm-cs}, \eqref{fterm-dil}, we observe that they are invariant under
the rescalings
\begin{equation}
\left\{\text{Im}(S),\text{Im}(Y^i),\text{Re}(\W),f_{1i},g_0^i\right\}\! \to\! \left\{ \alpha \text{Im}(S),\beta \text{Im}(Y^i), \beta^5\gamma \text{Re}(\W), \frac{\beta^{4}\gamma}{\alpha}f_{1i}, \beta^{3}\gamma g_0^i\right\}\!
\end{equation}
for arbitrary $\alpha$, $\beta$, $\gamma$. Without loss of generality, we can therefore choose
\begin{equation}
\text{Re}(\W)=\text{Im}(S)=\mathcal{A}\,. \label{scaling}
\end{equation}
Using this together with \eqref{ka}--\eqref{noscale1}, we find that \eqref{fterm-cs} is solved by
\begin{equation}
g_0^i = f_1^i\,, \qquad w_0 = 0\,, \label{g-cond}
\end{equation}
where $f_1^i=\K^{i\bar\jmath}f_{1\bar\jmath}$.
This is equivalent to the well-known ISD condition $F_3 = -\mathrm{e}^{-\phi}\star_6 H_3$ in 10d language \cite{Giddings:2001yu}.

The conditions that remain to be satisfied are then \eqref{wphi3} and \eqref{fterm-dil}. Using the above results, they simplify to the system of equations
\begin{equation}
\kappa_{ijk}f_1^jx^k = 0\,, \quad f_{1i}x^i=0\,,\quad f_{1i}\K^i = i\,. \label{key-eq1}
\end{equation}
Note that this can equivalently be written as
\begin{equation}
\kappa_{ijk}f_1^jx^k = 0\,, \quad \mathcal{A}_{i}x^i=0\,,\quad f_{1i}\K^i = i\,, \label{key-eq}
\end{equation}
as follows from contracting the first equation with $\K^i$ and then using \eqref{ka}--\eqref{noscale1}.

Eqs.~\eqref{key-eq1} are the main equations in this subsection.
We want to find a solution to this system for $x^i$ such that a long winding trajectory is obtained. We will argue that, if a sufficient number of axions mix, $x^i$ can be rotated into an aligned direction without a backreaction effect on the moduli. The overall normalisation of $x^i$ is irrelevant to show this and hence we will only consider the unit vector $\hat x^i$ from now on.

A key point for the following discussion is that the general solution $\hat x^i$ to \eqref{key-eq1} has free parameters. These parameters come in two types.
First, since the equations in \eqref{key-eq1} depend on $\K^i$ and $\K^{i\bar\jmath}$ (through $f_1^i=\K^{i\bar\jmath}f_{1\bar\jmath}$), the solution $\hat x^i$ will depend on the moduli vevs $\text{Im}(Y^i)$. Recall that those are unconstrained parameters since we already solved the corresponding $F$-term constraints by eliminating some of the fluxes (cf.\ \eqref{g-cond}). The $\text{Im}(Y^i)$ may therefore be set to any desired value compatible with our assumption of large complex structure.

Second, the solution $\hat x^i$ will depend on the flux numbers $f_{1i}$. As discussed earlier, $\hat x^i$ is fully determined by \eqref{key-eq1} if $f_{1i}$ is chosen such that the $n\times (n+1)$ matrix $(\kappa_{ijk}f_1^k,f_{1j})$ has rank $n-1$. Apart from this requirement, $f_{1i}$ is an arbitrary real vector with $n$ components that we may choose however we like. Let us also assume that we have a mixing of $p\le n$ axions, i.e., $p$ components of $x^i$ are non-zero. This implies that the unit vector $\hat x^i$ has $p-1$ independent components. For fixed moduli vevs, \eqref{key-eq1} thus depends on $n+p-1$ variables ($f_{1i}$ and $\hat x^i$) and yields at most $n+2$ linearly independent equations.  A sufficient condition for free flux parameters is therefore $n \ge p \ge 4$.\footnote{For our purposes, we need the existence of {\it real} solutions of our system of equations for the fluxes and the $\hat x^i$. Since that system (defined by the integers $\kappa_{ijk}$) is non-linear, this may impose extra conditions on the triple intersection numbers. We leave the study of this to future work.} We will see below that requiring $\hat x^i$ to depend on these parameters yields a further condition on the dual triple intersection numbers $\kappa_{ijk}$.
Let us denote the free flux parameters by $\lambda^i$.
The general solution to \eqref{key-eq1} is therefore of the form
\begin{equation}
\hat x^i = \hat x^i\!\left(\text{Im}(Y^i), \lambda^i\right)\,. \label{x-sol}
\end{equation}

We can now try to construct aligned or misaligned winding trajectories in this setting. Recall that misalignment means that $\hat x^i$ is aligned with a diagonal in the $p$-dimensional axion field space. On the other hand, alignment means that $\hat x^i$ is aligned with a hyperplane. We observe that (mis-)aligned winding trajectories can be engineered in two qualitatively different ways:
\begin{itemize}
\item The first option is to adjust the moduli vevs $\text{Im}(Y^i)$ such that $\hat x^i$ is (mis-)aligned. Since both $\hat x^i$ and $\mathcal{A}_{i\bar\jmath}/\mathcal{A}$ depend non-trivially on the moduli, it is difficult to judge whether this leads to large field ranges unless one sets out to perform a detailed model-by-model analysis. In particular, any change in $\hat x^i$ achieved by adjusting the $\text{Im}(Y^i)$ will in general backreact on $\mathcal{A}_{i\bar\jmath}/\mathcal{A}$ and thus potentially destroy the long trajectory. Indeed, we showed this backreaction to forbid large field ranges whenever $\hat x^i$ is aligned with a coordinate axis in the axion field space, cf.\ Sects.~\ref{sec:No_Go_Two_Axions_IIB} and \ref{sec:ThreeAxions}. In particular, this fully excluded aligned winding in the case $p=2$. On the other hand, we argued that, if $\hat x^i$ is aligned with a diagonal or a hyperplane, the backreaction does not generate large hierarchies in the moduli vevs such that our no-go theorem can be evaded.
\item The second option is to adjust the $\lambda^i$ parameters. Remarkably, they only appear in $\hat x^i$ but not in $\mathcal{A}_{i\bar\jmath}/\mathcal{A}$ such that making $\hat x^i$ large this way does \emph{not} backreact on the $\mathcal{A}_{i\bar\jmath}/\mathcal{A}$ factor.
It is therefore straightforward to determine when long trajectories can be realised, even without analysing particular models. As we will see below, $\lambda^i$ parameters arise if the number of mixing axions $p$ is large enough and a geometric condition involving the dual triple intersection numbers $\kappa_{ijk}$ is satisfied. The problem of constructing large field ranges is thus reduced to a condition purely on the geometry of the manifold.
\end{itemize}

We now discuss this second option in more detail. Without loss of generality, let us choose a basis such that $\hat x^i$ lies in the ``1'' direction and $f_{1i}$ in the ``2'' direction. Crucially, this is not necessarily a basis of the Kahler  cone. In this basis, \eqref{key-eq1} can be satisfied if
\begin{equation}
\kappa_{i1}{}^{\bar 2}=0\,, \quad \K^2 \neq 0\,. \label{def-0th-order1}
\end{equation}
The requirement that the direction of $\hat x^i$ is completely fixed by \eqref{key-eq1} for a given flux choice $f_{1i}$ (i.e., that $\text{rk}(\kappa_{ijk}f_1^k,f_{1j})=n-1$) becomes
\begin{equation}
\text{rk} (\kappa_{i j}{}^{\bar 2}, \delta_j^2) = n-1\,. \label{def-0th-order2}
\end{equation}
We now claim that a sufficient condition for free parameters $\lambda^i$ is
\begin{equation}
n \ge p \ge 4\,, \qquad n-p+1 < \text{rk} (\kappa_{1 j}{}^{\bar \alpha})\,, \label{def-cond}
\end{equation}
where $\alpha=3,\ldots,n$ labels the directions orthogonal to $\hat x^i$ and $f_{1i}$. To see this, consider a small deformation $f_{1i}\to f_{1i} + \epsilon_i$, $\hat x^i \to \hat x^i+\lambda^i(\epsilon_j)$ of a given solution to \eqref{key-eq1}. We argued above that there must be at least $p-3$ such deformations  (corresponding to $p-3$ free flux parameters). However, we have not excluded yet that these deformations leave $\hat x^i$ invariant, i.e., that $\epsilon_i\neq 0$ while $\lambda^i= 0$.
To show that this is not the case, let us assume that all $p-3$ deformations satisfy $\lambda^i=0$. Expanding \eqref{key-eq1} up to linear order and using \eqref{def-0th-order1}, we then find
\begin{equation}
\epsilon_1=0\,, \quad \epsilon_2 = -\frac{\epsilon_\alpha \K^\alpha}{\K^2}\,, \quad \kappa_{1j}{}^{\bar \alpha} \epsilon_{\bar \alpha} = 0\,, \label{def-cond2}
\end{equation}
where again $\alpha=3,\ldots,n$ and $\epsilon_{\bar \imath}=\epsilon_i$ because $\epsilon_i$ is real.
The last condition yields $n$ homogeneous equations for the $n-2$ components $\epsilon_\alpha$. If \eqref{def-cond} holds, at most $p-4$ of these components remain unfixed by \eqref{def-cond2}. Since the total number of linearly independent deformations is at least $p-3$, it then follows that there must be at least one deformation that is not captured by the ansatz $\lambda^i=0$. This proves our above claim that \eqref{def-cond} is a sufficient condition for the existence of free parameters $\lambda^i$.\footnote{Using similar reasoning, one may attempt to derive a (more complicated) condition which is both necessary \emph{and} sufficient. To keep the discussion simple, we omit a detailed derivation of such a refined condition here.}
Note, however, that this does not yet determine \emph{how} $\hat x^i$ depends on these parameters, i.e., how close to a hyperplane in the axion field space we can rotate $\hat x^i$ in a given model. It would therefore clearly be important to study our idea further on explicit Calabi-Yaus.

To summarise, we have argued that winding trajectories in Calabi-Yau compactifications at large complex structure are governed by the simple set of equations \eqref{key-eq1}. These admit a solution on any manifold for which \eqref{def-0th-order1}, \eqref{def-0th-order2} hold in some basis. The trajectories can be made long in two ways: either by adjusting the moduli vevs $\text{Im}(Y^i)$ or by adjusting the flux parameters $\lambda^i$. A sufficient condition for such parameters to exist is that the number of complex-structure moduli is at least 4 and a condition on the rank of the dual triple intersection numbers is satisfied, cf.\ \eqref{def-cond}. The problem of realising long trajectories in this setting thus reduces to a condition purely on the geometry.\footnote{Note that the conditions involving the triple intersection numbers in \eqref{def-0th-order1}--\eqref{def-cond} are in general not topological due to their dependence on inverse metric factors $\cK^{i\bar\jmath}$ and hence on the complex-structure moduli.} It will be interesting to study in detail whether there are indeed Calabi-Yaus satisfying this condition.

\begin{figure}[t]
\centering
 \includegraphics[width=0.8\textwidth]{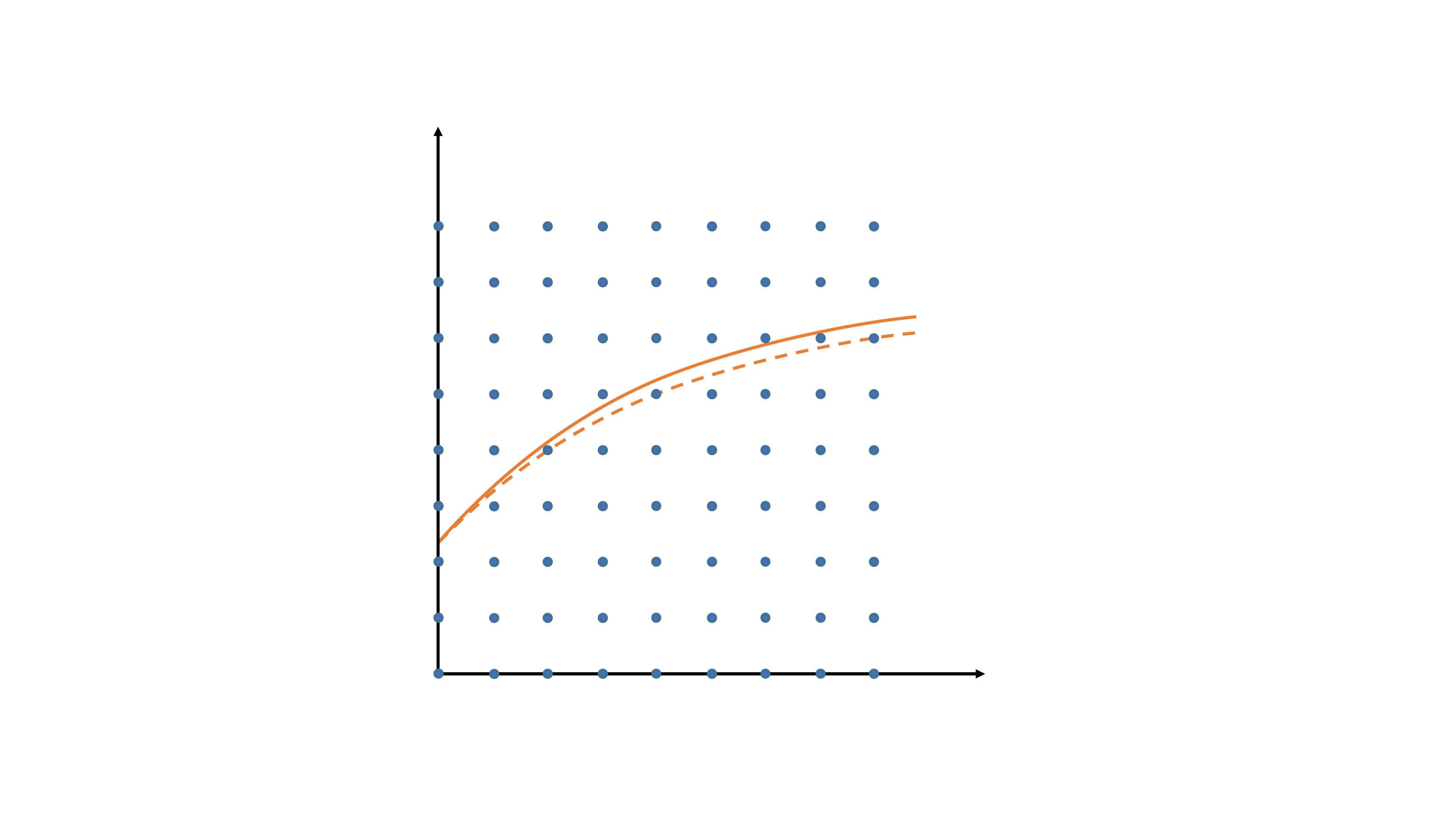}
 \put(-130,122){\footnotesize{$\epsilon(\lambda)$}}
 \put(-135,24){\footnotesize{$f_{11}$}}
 \put(-270,158){\footnotesize{$f_{12}$}}
\caption{2d slice of the flux lattice and the surface traced out by the free parameters $\lambda^i$ before and after adjusting the moduli vevs.}\label{fig:lambda}
\end{figure}

Finally, let us discuss two possible obstacles to the realisation of the above ideas on a concrete Calabi-Yau. First, we stress that the fluxes $f_{1i}$ cannot be made arbitrarily large but are bounded by tadpole cancellation through $|f_{1i}|^2 \sim Q^\text{loc}_{3}$, where $Q^\text{loc}_{3}$ denotes the combined D3-charge of the O3/O7-planes and D3/D7-branes in the compactification. Since aligning or misaligning the effective axion trajectory relies on large flux numbers, this implies that there is an upper bound on the possible enhancement of $f^2$. However, Calabi-Yaus can in general have rather large tadpoles such that we do not expect this to be a serious issue.
Second, we need to properly take into account flux quantisation. In order to simplify the discussion, we assumed above that the components of the flux vector $f_{1i}$ can take any real value while they are actually constrained to be integer. A possible concern is therefore that the surface traced out by the $\lambda^i$ parameters in flux space does not hit any points on the integer flux lattice (cf.\ Fig.\ \ref{fig:lambda}). The freedom to align $\hat x^i$ by adjusting the $\lambda^i$ parameters would then only be an artifact of our assumption of a real flux vector. While we do not present a detailed analysis here, it is plausible that this is not an issue for the following reason. As explained above, we still have the freedom to adjust the moduli vevs $\text{Im}(Y^i)$ however we like. Furthermore, we can use the shift symmetry of $\W$ to shift the non-integer parts of some of the flux numbers into the axion vevs $\text{Re}(Y^i)$. We expect that this freedom in $\text{Im}(Y^i)$, $\text{Re}(Y^i)$ can be used to slightly ``wiggle'' around the location $(f_{1i},g_0^j)$ of a solution in flux space and thus move it to a nearby properly-quantised point $(\tilde f_{1i},\tilde g_0^j)$ on the flux lattice. It will be interesting to study this in more detail in explicit constructions realising our idea.

\section{Effective field theory of the light axion}\label{sec:EFT_LF_AXION}

In this section, we discuss the stabilisation of the Kahler moduli in the presence of the light axion $\chi$. In particular, we work out the necessary conditions to ensure the required mass hierarchy for a low-energy effective field theory (EFT) for $\chi$.
While Sect.~\ref{sec:mass-scales} is devoted to summarising our most important results, Sect.~\ref{sec:Corrections} shows in detail that tree-level and loop corrections to the moduli masses and to the potential for $\chi$ can consistently be neglected at sufficiently large volume and small string coupling. We also investigate the possible role of a complex-structure dependence of the non-perturbative effects in the superpotential.
In Sect.\ \ref{sec:Towards_Inflation}, we analyse whether models of large-field inflation can be realised within the winding scenario. We find this to be challenging due to additional phenomenological constraints, which are in tension with the previous requirement of large volume.

\subsection{Mass scales and axion potential}
\label{sec:mass-scales} 

As we will see below, consistency of the EFT requires $\cV\gg 1$. We therefore stabilise the Kahler moduli according to the large-volume scenario (LVS) \cite{8, Conlon:2005ki}. Before we address our setup with one light axion, let us briefly recall the original LVS setup where all complex-structure moduli are stabilised by fluxes. For simplicity, we will focus on the simple example $\mathbb{CP}^4_{(1,1,1,6,9)}[18]$, which has only two Kahler moduli. The volume\footnote{We use a notation where the volume of the Calabi-Yau is given by $\cV=\frac{1}{6}\int_{X_3}\, J\wedge J\wedge J=\frac{1}{6}\, k_{ABC}\, t^{A}t^{B}t^{C}$ in terms of basis $2$-cycle volumes $t^{A}$. We define the complexified Kahler moduli as $T_{A}=b_{A}+i\tau_{A} $, where the $4$-cycle volumes $\tau_{A}$ are related to the $t^{B}$ as $\tau_{A}=\partial_{t^{A}}\cV$.} in terms of the $4$-cycle volumes $\tau_{A}$ is then of swiss-cheese type, i.e.,
\begin{equation}
\cV= \frac{1}{9\sqrt{2}}(\tau_{b}^{{3}/{2}}-\tau_{s}^{{3}/{2}})\,.
\end{equation}
The $4$-cycle volume $\tau_{b}$ controls the size of the Calabi-Yau, while $\tau_{s}$ parametrises the size of a small blow-up cycle in the Calabi-Yau. After having fixed all complex-structure moduli via fluxes, the superpotential including non-perturbative corrections is of the form
\begin{equation}\label{eq:WKMLVS} 
\cW=\cW_{0}+A_{s}\, \mathrm{e}^{ia_{s}T_{s}}\, ,
\end{equation}
where $\W_0$ denotes the vev of the tree-level flux superpotential. To next-to-leading order, the $\alpha^{\prime}$-corrected Kahler potential is
\begin{equation}\label{eq:KaehlerPotWithAP} 
\cK = \cK_\text{cs} - \log \frac{2}{g_s} -2\log\left (\cV+\dfrac{\xi}{2g_{s}^{\frac{3}{2}}}\right )\, ,
\end{equation}
where $\xi=-{\chi(X_{3})\zeta(3)}/{2(2\pi)^{3}}$ and $\chi(X_{3})$ is the Euler characteristic of the Calabi-Yau threefold $X_{3}$. This leads to a scalar potential of the form
\begin{equation}\label{eq:PotKM} 
V_\text{LVS}(\tau_{s},\tau_{b})=g_{s}\mathrm{e}^{\cK_\text{cs}}\left (\dfrac{12\sqrt{2 \tau_{s}}}{\cV}(a_{s}A_{s})^{2}\,\mathrm{e}^{-2a_{s}\tau_{s}}-\dfrac{2|\cW_{0}|\tau_{s}}{\cV^{2}}\, a_{s}|A_{s}|\, \mathrm{e}^{-a_{s}\tau_{s}}+\dfrac{3}{8}\dfrac{\xi|\cW_{0}|^{2}}{g_{s}^{\frac{3}{2}}\cV^{3}}\right )\,,
\end{equation}
where the axionic partner of $\tau_s$ has already been stabilised. At the LVS-minimum, we have \cite{8}
\begin{equation}\label{eq:LVSMin} 
\tau_{s}\sim \dfrac{\xi^{\frac{2}{3}}}{g_{s}}\,, \qquad \cV\sim \dfrac{ |\cW_{0}|}{a_{s}|A_{s}|\sqrt{g_{s}}}\, \xi^{\frac{1}{3}}\mathrm{e}^{a_{s}\tau_{s}}\,.
\end{equation}
The on-shell value of the scalar potential (and, hence, the AdS curvature scale) is given by (cf.\ Eq.~(B.18) in \cite{35})
\begin{equation}\label{eq:LVSMinPot} 
V_\text{LVS}\bigl |_{\tau_{s},\cV} \sim -m_\text{AdS}^2 \sim -\dfrac{\sqrt{g_{s}}\mathrm{e}^{\cK_\text{cs}}\xi^{\frac{1}{3}}|\cW_{0}|^{2}}{\cV^{3}}\, .
\end{equation}
In the large-volume limit, the modulus $\tau_{s}$ is heavy in comparison to the volume modulus $\tau_{b}$ \cite{Conlon:2005ki}:
\begin{align}\label{eq:MassTauB} 
m_{\tau_{s}}\sim\mathrm{e}^{\cK_\text{cs}/2}\dfrac{|\cW_{0}|{\xi}^{2/3}}{\sqrt{g_{s}}\cV}\, , \qquad m_{\tau_{b}}\sim\mathrm{e}^{\cK_\text{cs}/2}\dfrac{|\cW_{0}|\sqrt{\xi}}{g_{s}^{1/4}\cV^{3/2}}\, .
\end{align}
For completeness, we note that the axion associated to $\tau_s$ is stabilised at the same scale as $\tau_s$. The axion associated to $\tau_b$ is effectively massless as it only receives a mass $\sim\mathrm{e}^{-\tau_{b}}$ from non-perturbative effects neglected in \eqref{eq:WKMLVS}, \eqref{eq:KaehlerPotWithAP}. This axion has a tiny decay constant $f_{b_{b}}\sim \cV^{-2/3}\ll 1$  \cite{Cicoli:2012sz}.
We therefore have an additional light axion of small field range present in our EFT. This does not pose a problem for Sect.~\ref{sec:Towards_Inflation} because, in the large-volume limit, the axion is so light that it will play no role during the inflationary dynamics \cite{Conlon:2005ki,33}. 

After having reviewed the LVS, we now focus on the winding scenario, where one complex-structure axion remains unstabilised by the fluxes. As in Sect.~\ref{sec:General_Argument}, we denote by $X$ the complex field whose real part is the light axion $\chi$. As before, $X$ is some linear combination of a subset $U_i$ of the complex-structure moduli $Y^i = \left\{ U_i, Z \right\}$ with $U_i=x_iX$ and $x_i$ some integer numbers. Recall that the axionic shift symmetry of $\chi$ was ensured in the previous sections by working at large complex structure for the moduli $U_{i}$.
This shift symmetry is broken classically by terms $\sim \mathrm{e}^{i U_{i}}$, leading to a periodic potential for $\chi$ \cite{Hebecker:2015rya}.

In order to derive the potential for $\chi$, we consider
\begin{equation}\label{eq:FullPotential} 
V=\mathrm{e}^{\cK}\left (\cK^{T_{A}\bar{T}_{B}}\, D_{T_{A}}\cW\, D_{\bar{T}_{B}}\overline{\cW}+\cK^{X\bar{X}}\, \bigl |D_{X}\cW\bigl |^{2}-3|\cW|^{2}\right )\,,
\end{equation}
where we have already integrated out the axio-dilaton and all complex-structure moduli apart from $X$. Let us denote the individual parts of the scalar potential by
\begin{equation}
V_\text{GKP}=\mathrm{e}^{\cK}\cK^{X\bar{X}}\, \bigl |D_{X}\cW\bigl |^{2}\,, \qquad V_\text{LVS}=\mathrm{e}^{\cK}\left (\cK^{T_{A}\bar{T}_{B}}\, D_{T_{A}}\cW\, D_{\bar{T}_{B}}\overline{\cW}-3|\cW|^{2}\right )\,
\end{equation}
so that the full potential can be written as $V=V_\text{LVS}+V_\text{GKP}$. Here, $V_\text{GKP}$ is assumed to generate the leading potential for $\chi$, while $V_\text{LVS}$ yields the familiar potential \eqref{eq:PotKM} stabilising the Kahler moduli. We will show in Sect.~\ref{sec:Corrections} that this assumption is self-consistent at sufficiently large $\mathcal{V}$ and small $g_s$.

The superpotential and Kahler potential are again given by \eqref{eq:WKMLVS} and \eqref{eq:KaehlerPotWithAP}.
For convenience, we split $\cW_0$ and $\cK_\text{cs}$ into a large-complex-structure part, which is independent of $\chi$, and an exponentially suppressed part, which produces the potential for $\chi$:
\begin{align}
& \cW_0(X)=\cW_\text{lcs}+\cW_\text{ax}(X)\,, && \cK_\text{cs}(X,\bar X)=\cK_\text{lcs}(X-\bar X)+\cK_\text{ax}(X,\bar X)\,, \notag \\
& \cW_\text{ax}\sim \mathrm{e}^{iU_i} \sim \mathrm{e}^{-u_{i}}\,, && \cK_\text{ax}\sim \frac{i\left(U_i-\bar U_i\right)}{\mathcal{A}}\left(\mathrm{e}^{iU_i}+\mathrm{e}^{-i\bar U_i}\right) \sim \frac{u_{i}\mathrm{e}^{-u_{i}}}{\cA}\,. \label{k_ax}
\end{align}
Here, we have only schematically displayed the $U_i$-dependence and the orders of magnitude of $\cW_\text{ax}$ and $\cK_\text{ax}$. We refer to \cite{Hebecker:2015rya} for fully explicit expressions.\footnote{There are in general also terms $\cW_\text{ax}\sim u_i\mathrm{e}^{-u_{i}} \gg \mathrm{e}^{-u_{i}}$ but these have to be put to zero by an appropriate flux choice in order to allow the structure \eqref{eq:W1.1} \cite{Hebecker:2015rya}.}

If we work in a regime where $A_{s}\sim\cO(1)$ and $\mathrm{e}^{-u_i}\gg \mathrm{e}^{-a_s\tau_s}$, we have $\partial_X \cW \approx \partial_X \cW_\text{ax}$. Let us also assume $\p_{X}\cW_\text{ax}\ll \cK_{X}\cW$, e.g., by choosing $\cW_{0}$ sufficiently large. We will again see in Sect.~\ref{sec:Corrections} that these assumptions are self-consistent. Recall furthermore that $D_X \cW=0$ implies $\cK_X =0$ in the large-complex-structure limit for all values of $\chi$. Including the exponentially suppressed corrections \eqref{k_ax}, we can therefore estimate $\cK_X \sim ix_\star u_\star\mathrm{e}^{-u_{\star}}/\cA$, where we used $\partial U_i/\partial X = x_i$ and $u_\star$ denotes the smallest of the $u_i$. Crucially, this estimate for $\cK_X$ holds off-shell, i.e., away from the minimum for $\chi$. Using this, the leading contribution to the potential for $\chi$ is
\begin{align} \label{chi-pot}
V_\text{GKP}&\sim \dfrac{g_{s}\mathrm{e}^{\cK_\text{cs}}}{\cV^{2}}\, |\cW_{0}|^{2}\, \cK^{X\bar{X}} \cK_{X}\cK_{\bar{X}} \sim\dfrac{g_{s}\mathrm{e}^{\cK_\text{cs}}}{\cV^{2}}\, \frac{|\cW_{0}|^{2}}{\cA^2}\, \cK^{X\bar{X}} x_\star^2 u_\star^2 \mathrm{e}^{-2u_\star} |\lambda(x_\star\chi)|^2\,,
\end{align}
where $\lambda$ is a complex periodic function of $\chi$. At the minimum, we have $\cK_X \approx 0$ and, hence, $\lambda\approx 0$.

The axion mass is given by $m_\chi^2 \sim \partial_\chi^2 V_\text{GKP}/\cK_{X\bar{X}}$. Hence,
\begin{equation}
m_{\chi} \sim\mathrm{e}^{\cK_\text{cs}/2}\, \dfrac{\sqrt{g_{s}} |\cW_{0}|}{\cV\cA}\, \dfrac{x_\star^2 u_\star\mathrm{e}^{-u_\star}}{f^{2}}\, , \label{chim}
\end{equation}
where we used that $\cK^{X\bar{X}}\gtrsim \cK_{X\bar{X}}^{-1}=f^{-2}$ by Eq.~\eqref{fieldrange1}.
In order to keep track of the different instanton contributions to this mass, we can associate different mass scales to them. Let us denote the mass scale associated to the instanton with the longest periodicity by $m_{\chi}^\text{long}$ and the mass scales associated to the shorter instantons collectively by $m_{\chi}^\text{short}$. This distinction will in particular become important in the context of inflation, where the potential generated by the long instanton must dominate, see Sect.~\ref{sec:Towards_Inflation}.
We thus find
\begin{equation}\label{eq:ShortLongMass} 
m_{\chi}^\text{long}\sim\mathrm{e}^{\cK_\text{cs}/2}\, \dfrac{\sqrt{g_{s}} |\cW_{0}|}{\cV\cA}\, \dfrac{u_\text{long}\mathrm{e}^{-u_\text{long}}}{f^{2}}\,,  \qquad m_{\chi}^\text{short}\sim\mathrm{e}^{\cK_\text{cs}/2}\, \dfrac{\sqrt{g_{s}} |\cW_{0}|}{\cV\cA}\, \dfrac{N^2 u_\text{short}\mathrm{e}^{-u_\text{short}}}{f^{2}}\, .
\end{equation}
Here, we used that, for the instanton with the longest periodicity, we have $x_{i}\sim \cO(1)$, while the short instantons have $x_i\sim\cO(N)$.

In order to consistently integrate out $\tau_b$ and arrive at an EFT for the light axion $\chi$, we need to establish a hierarchy in the mass scales such that
\begin{equation}\label{eq:HierarchySLT} 
m_{\chi}^\text{long},\, m_{\chi}^\text{short}\ll m_{\tau_{b}}\, .
\end{equation}
Comparing \eqref{eq:MassTauB} to \eqref{eq:ShortLongMass} shows that this poses no big problem as the $u_i$ vevs can easily be tuned and the dependence on them is exponential.

\subsection{Corrections}\label{sec:Corrections}

As stated above, we still need to show that there is a regime where our stabilisation scheme is self-consistent. In particular, we assumed that the Kahler moduli are stabilised by the $V_\text{LVS}$-part of the scalar potential (as in the usual LVS), while the potential for $\chi$ is generated by the terms in $V_\text{GKP}$. This requires that corrections to the $\chi$-potential from $V_\text{LVS}$ as well as corrections to the $\tau_A$ masses from $V_\text{GKP}$ are subleading. Furthermore, there can be loop corrections to the scalar potential, which need to be subleading as well. We have summarised the different mass scales in our scenario and the magnitudes of the various tree-level and loop corrections in Table~\ref{tab:1}. As we will see below, all corrections can self-consistently be neglected in the regime
\begin{equation}
\dfrac{Nf}{x_\star\sqrt{\cV}g_{s}^{1/4}} \ll \frac{x_{\star}u_{\star}\mathrm{e}^{-u_{\star}}}{\cA}\ll \dfrac{f}{\sqrt{\cV}g_{s}^{3/4}}\, ,
\end{equation}
which is satisfied for, e.g.,
\begin{equation}
\frac{x_\star u_\star\mathrm{e}^{-u_\star}}{\cA} \sim \frac{\sqrt{N}f}{\sqrt{x_\star g_s\cV}}\,, \qquad g_s \ll \frac{x_\star^2}{N^2}\,. \label{regime}
\end{equation}
In the same regime, the AdS curvature scale is small compared to the saxion masses, setting our type IIB constructions apart from the type IIA model of Sect.~\ref{sec:axalCY}.\footnote{One furthermore checks that $V_\text{GKP}\gg V_\text{LVS}$ in this regime such that the total vacuum energy is positive, except very close to the minimum of the $\chi$-potential.}
We will also see that the above constraints on $u_\star$ and $g_s$ are relaxed significantly if $A_s$ in \eqref{eq:WKMLVS} can be assumed to be independent of $\chi$.
Readers not interested in the detailed derivation of these results may skip directly to Sect.~\ref{sec:Towards_Inflation}.

\subsubsection{Tree-level corrections to the Kahler moduli masses}

Let us begin with the term $V_\text{GKP}$, which yields corrections $\delta m_{\tau_{A}}^\text{GKP}$ to the Kahler moduli masses. We find that\footnote{Here and in the following, we define $\delta m$ as the square-root of the correction to $m^2$, i.e., the total squared mass is given by $m^2+(\delta m)^2$ rather than $(m+\delta m)^2$.}
\begin{equation}
\left (\delta m_{\tau_{A}}^\text{GKP}\right )^{2}=\dfrac{\p_{\tau_{A}}^{2}V_\text{GKP}}{\cK_{T_{A}\bar{T}_{A}}}\sim \mathrm{e}^{\cK_\text{cs}}\dfrac{g_{s}}{\sqrt{\tau_{A}}\cV^{3}}\dfrac{\cK^{X\bar{X}}}{\cK_{T_{A}\bar{T}_{A}}}\, \cK_{X}\cK_{\bar{X}} |\cW_{0}|^{2}\,.
\end{equation}
According to the discussion above \eqref{chi-pot}, we have
$\cK_{X}\sim ix_{\star}u_{\star}\mathrm{e}^{-u_{\star}}/\cA$. Since $\cK_{T_{A}\bar{T}_{A}}\sim (\cV\sqrt{\tau_{A}})^{-1}$ and $\cK^{X\bar{X}}\gtrsim f^{-2}$, this amounts to
\begin{equation}\label{eq:GKPCorTau} 
\left (\delta m_{\tau_{A}}^\text{GKP}\right )^{2}\sim \mathrm{e}^{\cK_\text{cs}}\dfrac{g_{s}}{\cV^{2}}\, \dfrac{|\cW_{0}|^{2}x_{\star}^{2}u_{\star}^{2}\mathrm{e}^{-2u_{\star}}}{\cA^2 f^{2}}\, .
\end{equation}
We want to enforce the condition
\begin{equation}\label{eq:GKPCorTauHie} 
m_{\tau_{A}}\gg \delta m_{\tau_{A}}^\text{GKP}
\end{equation}
so that the Kahler moduli are not in danger of being destabilised by the corrections. Comparing \eqref{eq:GKPCorTau} with \eqref{eq:MassTauB}, we conclude that this can always be achieved by a suitable tuning of the saxion values $u_{i}$. Explicitly, we find for $\tau_{b}$ the necessary requirement
\begin{equation}\label{eq:GKPCorTauHieCons} 
\frac{x_{\star}u_{\star}\mathrm{e}^{-u_{\star}}}{\cA}\ll \dfrac{f}{\sqrt{\cV}g_{s}^{3/4}}\, .
\end{equation}
One checks that this inequality is indeed satisfied in the regime \eqref{regime}. Also note that it follows from \eqref{chim} that $\delta m_{\tau_{A}}^\text{GKP}\sim \frac{f}{x_\star}m_{\chi}\gtrsim m_{\chi}$, where we assumed $f\sim \mathcal{O}(N)$ and $x_\star \lesssim \mathcal{O}(N)$. Hence, \eqref{eq:GKPCorTauHie} already implies \eqref{eq:HierarchySLT}.

\begin{table}[t]
\centering
\begin{tabular}{|c|c|c|}
\hline 
\rule[-1ex]{0pt}{4.5ex} Mass scales & $A_{s}=\text{const.}$ & $A_{s}=A_{s}(\chi)$ \\ [0.5em]
\hline 
\rule[-1ex]{0pt}{5ex} $m_{\tau_{s}}$ & \multicolumn{2}{c|}{$\dfrac{|\cW_{0}|}{\sqrt{g_{s}}\cV}$} \\[1em]
\hline 
\rule[-1ex]{0pt}{5ex} $m_{\tau_{b}}$ & \multicolumn{2}{c|}{$\dfrac{|\cW_{0}|}{g_{s}^{1/4}\cV^{3/2}}$} \\[1em]
\hline 
\rule[-1ex]{0pt}{5ex} $\delta m_{\tau_{b}}^\text{GKP}$ & \multicolumn{2}{c|}{$ \dfrac{\sqrt{g_{s}}\, |\cW_{0}|}{\cV}\, \dfrac{x_\star u_{\star}\mathrm{e}^{-u_{\star}}}{\cA f}$} \\[1em]
\hline 
\rule[-1ex]{0pt}{5ex} $m_{\chi}$ & \multicolumn{2}{c|}{$\dfrac{\sqrt{g_{s}}\, |\W_0|}{\cV}\, \dfrac{x_\star^2 u_{\star} \mathrm{e}^{-u_{\star}}}{\cA f^{2}}$} \\[1em]
\hline 
\rule[-1ex]{0pt}{5ex} $m_\text{AdS}$ & \multicolumn{2}{c|}{$\dfrac{g_{s}^{1/4}|\cW_{0}|}{\cV^{3/2}}$} \\[1em]
\hline
\rule[-1ex]{0pt}{5.5ex} $\delta m_{\chi}^\text{LVS}$ & $ \dfrac{g_{s}^{{1}/{4}}|\cW_{0}|}{\cV^{{3}/{2}}}\, \dfrac{x_{\star}\sqrt{u_{\star}}\mathrm{e}^{-u_{\star}/2}}{\sqrt{\cA} f}$ & $ \;\qquad\dfrac{g_{s}^{{1}/{4}}|\cW_0|}{\cV^{{3}/{2}}}\, \dfrac{N}{f} \qquad\;$ \\[1em] 
\hline 
\rule[-1ex]{0pt}{5.5ex} $\delta m^\text{1-loop}_{\tau_{b}}$ & \multicolumn{2}{c|}{$\dfrac{g_{s}^{3/2}\, |\cW_{0}|}{\cV^{5/3}}$} \\[1em] 
\hline 
\rule[-1ex]{0pt}{5.5ex} $\delta m^\text{1-loop}_{\chi}$ & \multicolumn{2}{c|}{$ \dfrac{g_{s}^{3/2}\, |\cW_{0}|}{\cV^{5/3}}\, \dfrac{N}{f}$} \\[1em]
\hline 
\end{tabular} 
\caption{The table summarises the appearing mass scales and the magnitude of the different corrections.
We divided all results by a factor of $\mathrm{e}^{\cK_\text{cs}/2}$.}\label{tab:1} 
\end{table}

\subsubsection{Tree-level corrections to the axion potential}

Next, we consider possible corrections $\delta m_{\chi}^\text{LVS}$ to the $\chi$-potential from $V_\text{LVS}$. It turns out that the magnitude of these corrections depends crucially on whether $A_s$ in the superpotential \eqref{eq:WKMLVS}  is assumed to depend on $\chi$ or not.
For a general Calabi-Yau threefold, $A_{s}(Y^i)$ is an unknown function of the complex-structure moduli $Y^i$ \cite{Kachru:2003aw,Baumann:2006th}. Its explicit dependence has so far only been computed for simple examples such as toroidal $\cN=1$ orientifolds \cite{Berg:2004ek,Berg:2004sj}. However, one can argue that $A_s$ in our case is not a function of $\chi$. Before we discuss this conjecture in more detail, let us first state the corrections $\delta m_{\chi}^\text{LVS}$. We do this both for the case where $A_s$ is constant in $\chi$ and for the case where
it is a function of $\chi$ with $A_{s}\sim\p_{U_{i}}A_{s}\sim\cO(1)$.

Consider first the case $A_s=A_s(\chi)$. With \eqref{eq:LVSMin} and \eqref{eq:LVSMinPot}, we find that
\begin{equation}\label{eq:LVSChiA1}
\left (\delta m_{\chi}^\text{LVS}\right )^{2}=\dfrac{1}{\cK_{X\bar{X}}}\p_{\chi}^{2} V_\text{LVS}\bigl |_{\tau_{s},\cV}\,\sim\dfrac{N^{2}}{\cK_{X\bar{X}}} V_\text{LVS}\bigl |_{\tau_{s},\cV}\,\sim\mathrm{e}^{\cK_\text{cs}}\dfrac{\sqrt{g_{s}}|\cW_{0}|^{2}}{\cV^{3}}\, \dfrac{N^{2}}{f^{2}}\,,
\end{equation}
where we used that $\p_{\chi}A_{s}\sim x_{i}\cO(1)\lesssim N$ as well as $\cK_{X\bar{X}}=f^{2}$.
Requiring $\delta m_{\chi}^\text{LVS}\ll m_{\chi}$ yields
\begin{equation}
\frac{x_{\star}^2u_{\star}\mathrm{e}^{-u_{\star}}}{\cA}\gg \dfrac{Nf}{\sqrt{\cV}g_{s}^{1/4}}\, . \label{m-chi-lvs}
\end{equation}
To successfully achieve both \eqref{m-chi-lvs} and \eqref{eq:GKPCorTauHieCons}, we require a small string coupling, $g_{s}\ll x_\star^2/N^2$. In the regime \eqref{regime}, all inequalities are satisfied.

Consider now the case where $A_{s}$ is independent of $\chi$. The leading correction $\delta m_{\chi}^\text{LVS}$ is now due to the $\chi$-dependence of the prefactor $\mathrm{e}^{\cK_\text{cs}}$ in \eqref{eq:LVSMinPot}. Using $\p_{\chi}^{2}\cK_\text{cs}\sim x_{\star}^{2} u_{\star}\mathrm{e}^{-u_{\star}}/\cA$, we find
\begin{equation}\label{eq:LVSChiA2} 
\left (\delta m_{\chi}^\text{LVS}\right )^{2}=\dfrac{1}{\cK_{X\bar{X}}}\p_{\chi}^{2} V_\text{LVS}\bigl |_{\tau_{s},\cV}\sim\mathrm{e}^{\cK_\text{cs}}\dfrac{\sqrt{g_{s}}|\cW_{0}|^{2}}{\cV^{3}}\, \dfrac{x_{\star}^{2}u_{\star}\mathrm{e}^{-u_{\star}}}{\cA f^{2}}\, .
\end{equation}
This is exponentially suppressed by the extra factor $x_\star^2u_{\star}\mathrm{e}^{-u_{\star}}/(N^2\cA)$ in comparison to Eq.~\eqref{eq:LVSChiA1}. Imposing the hierarchy $\delta m_{\chi}^\text{LVS}\ll m_{\chi}$ now implies that 
\begin{equation}
\frac{x_\star^2 u_\star \mathrm{e}^{-u_{\star}}}{\cA}\gg \dfrac{f^2}{\sqrt{g_{s}}\cV}\, . \label{m-chi-lvs2}
\end{equation}
We observe that satisfying this together with \eqref{eq:GKPCorTauHieCons} is much less constraining than having to satisfy \eqref{m-chi-lvs} and \eqref{eq:GKPCorTauHieCons}.
In particular, if $A_s$ is independent of $\chi$, we can relax the condition $g_s\ll x_\star^2/N^2$. Note, however, that small $g_s$ also ensures that the AdS scale is small compared to $m_{\tau_b}$ (cf.~Table~\ref{tab:1}).

\begin{figure}
\centering
 \includegraphics[width=0.6\textwidth]{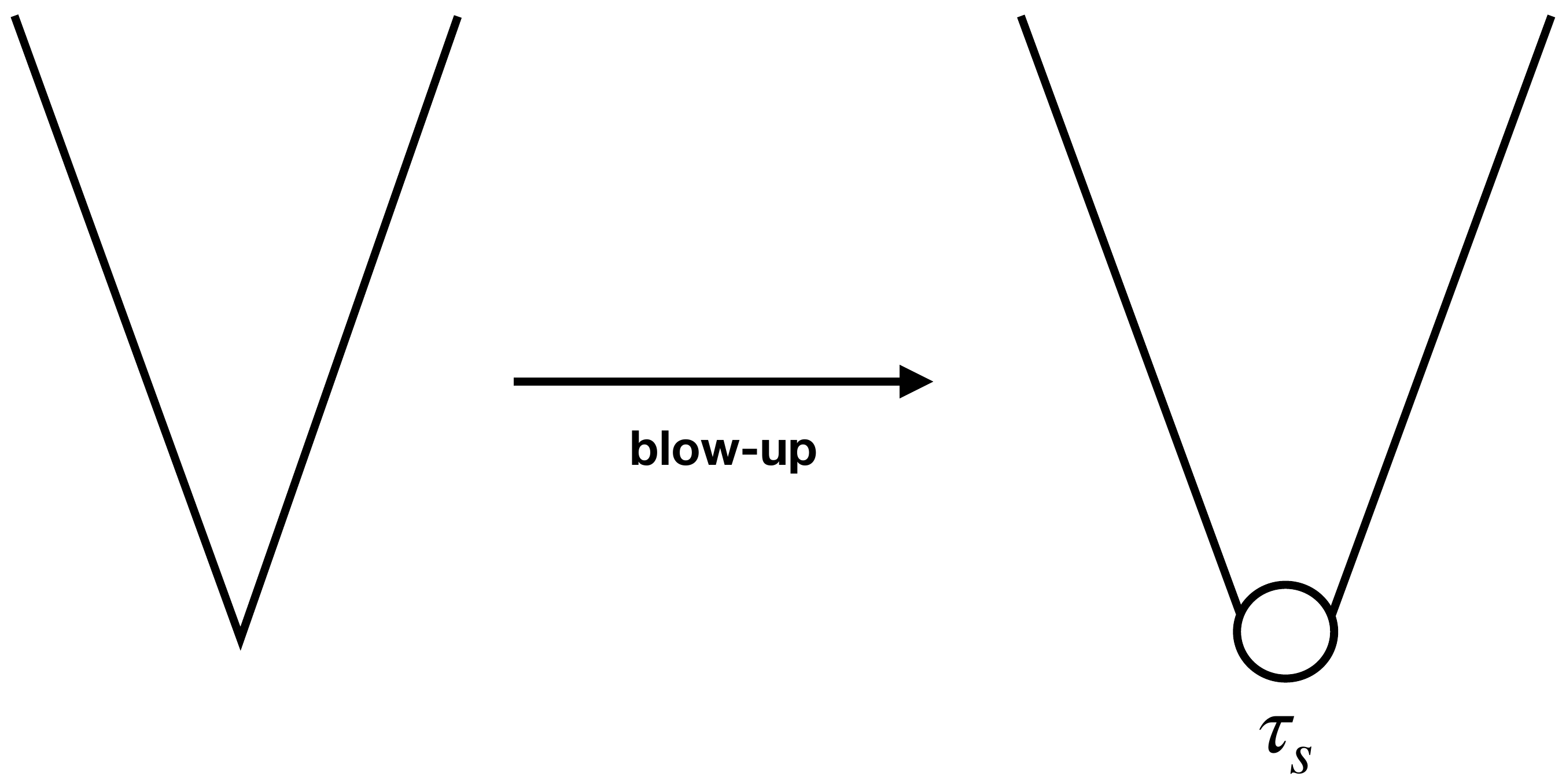}
\caption{The point-like singularity is resolved by the blow-up cycle $\tau_{s}$ by replacing the point with a projective space $\mathbb{C}\mathbb{P}^{1}$. The figure is adapted from \cite{20}.}\label{fig:5}
\end{figure}

Let us now motivate in more detail the possibility that $A_s$ is constant in $\chi$.
Recall that, in deriving the LVS minimum, it turns out to be crucial that $\tau_{s}$ is a blow-up mode of a point-like singularity \cite{8, 20} (cf.~Fig.~\ref{fig:5}).
A point-like singularity can be blown up by replacing it with a projective space like $\mathbb{C}\mathbb{P}^{1}$, thereby introducing an ``exceptional'' divisor. This divisor has an associated Kahler modulus, which in our case corresponds to $\tau_{s}$. Further studies \cite{Cicoli:2011it} showed that a natural candidate for a blow-up mode supporting the non-perturbative effect in $\cW$ is a so-called ``diagonal'' del Pezzo divisor. Such a blow-up is  local, holomorphic and leaves the complex structure invariant, cf., e.g., \cite{Grimm:2008ed} Sect.~$4.3.4$. Indeed, the 4-cycle parametrised by $\tau_s$ in our example is of this type. 
If the local geometry close to the singularity involves $3$-cycles, then there will be a backreaction of the complex structure on the blow-up. To avoid this, we assume that no such local 3-cycles are present.
In particular, consider cutting off the part of the Calabi-Yau containing the blow-up and taking a limit where the boundary is taken to infinity at fixed blow-up-cycle volume.
Our assumption is that, in such a non-compact limit, the blow-up does not possess any complex-structure deformations.
This guarantees
that, in the limit of small $\tau_{s}$, the complex-structure dependence of $A_{s}$ introduced by the D3-instanton on the shrinking cycle can be neglected.
Note that, even though $\tau_{s}\sim 1/g_{s}\gtrsim \mathcal{O}(1)$ at the LVS minimum, understanding the limit $\tau_{s}\raw 0$ is still relevant for our setup. Indeed, it is known that $A_s$ does not depend on the Kahler moduli. Hence, if $A_s$ can be argued to not depend on $\chi$ for small $\tau_s$, the same must be true at large $\tau_s$. We will leave a more careful treatment of a possible complex-structure dependence of $A_{s}$ for future works.

\subsubsection{Loop corrections}

Let us finally discuss loop corrections to the scalar potential.
While the superpotential only receives non-perturbative corrections, the Kahler potential can obtain contributions at every order in perturbation theory. In particular, there are loop corrections which depend on the complex-structure moduli \cite{21,17,18,19} and could therefore break the shift symmetry of $\chi$.
The Kahler potential also receives non-perturbative contributions from brane or worldsheet instantons, but they are subdominant in comparison to the perturbative corrections and will therefore be ignored in the following, see, e.g., \cite{15,16}.
The known loop corrections to the Kahler potential satisfy an extended no-scale structure such that they affect the scalar potential only at subleading order in the volume \cite{21,19,17,18, Junghans:2014zla}. In particular, the scalar potential for our example $\cV\sim (\tau_{b}^{{3}/{2}}-\tau_{s}^{{3}/{2}})$ receives a $1$-loop-correction \cite{19}
\begin{equation}\label{eq:LC:Est1L} 
V_\text{1-loop}\sim\mathrm{e}^{\cK_\text{cs}}\dfrac{|\W_{0}|^{2}}{\cV^{3}}  \left \{\dfrac{g_{s}^{3} \left(\cC_{b}^{K}(Y,\bar{Y})\right)^{2}}{\cV^{1/3}}+g_{s}^{\frac{3}{2}} \cC_{s}^{K}(Y,\bar{Y}) \right \}\,,
\end{equation}
where we used $\tau_s\sim 1/g_s$ and neglected $\mathcal{O}(1)$ prefactors as well as further $\cC_{b/s}^{K}$-dependent terms subleading in $g_s$. The correction is due to the exchange of Kaluza-Klein modes between D7-branes wrapped on the 4-cycles associated to $\tau_{b/s}$ and D3-branes localised on the internal manifold (or equivalently between O7- and O3-planes).\footnote{Notice that, for the example under consideration, there are no further contributions due to the exchange of winding modes since the two divisors do not intersect \cite{Curio:2006ea, 19}.} The coefficients $\cC_{b}^{K}$ and $\cC_{s}^{K}$ are functions of the complex-structure moduli whose explicit form is in general unknown.\footnote{For toroidal orientifolds, $\cC^{K}$ is given by Eisenstein series involving polynomial as well as exponential terms in the complex-structure moduli \cite{17,18}.}

We ensure that the second term $\sim g_{s}^{3/2}\cC_{s}^{K}$ vanishes by assuming that there is no D7-brane wrapped on $\tau_{s}$. Hence, the non-perturbative effects $\sim A_s \mathrm{e}^{-a_s\tau_s}$ in the superpotential have to be generated by D3-brane instantons. Using $(\delta m_{\chi}^\text{1-loop})^2 \sim \partial_\chi^2V_\text{1-loop}/\cK_{X\bar X}$, we then find that the loop effects contribute at a scale
\begin{equation}
\delta m_{\chi}^\text{1-loop}\sim \mathrm{e}^{\cK_\text{cs}/2}\dfrac{g_{s}^{\frac{3}{2}}|\cW_{0}|}{\cV^{\frac{5}{3}}}\, \dfrac{N}{f}\,.
\end{equation}
Here, we assumed that $\cC_{b}^{K}\sim\cO(1)$, $\p_{U_{i}}\cC_{b}^{K}\sim\cO(1)$, which implies $\p_{\chi}\cC_{b}^{K}\sim x_{i}\cO(1)\lesssim N\cO(1)$. We observe that the loop corrections are suppressed by an additional volume factor $\cV^{-1/6}$ compared to the tree-level corrections \eqref{eq:LVSChiA1}. Ensuring that the tree-level corrections are negligible thus implies that also the loop corrections can be neglected,
\begin{equation}
\delta m_{\chi}^\text{1-loop} \ll m_{\chi}\,.
\end{equation}
Similarly, as is well known, loop corrections to the Kahler moduli masses are suppressed by a volume factor and can therefore consistently be neglected as well, $\delta m_{\tau_{A}}^\text{1-loop} \ll m_{\tau_{A}}$.
We have thus shown that both tree-level and loop corrections to the moduli masses and the axion potential of Sect.~\ref{sec:mass-scales} are negligible in the regime \eqref{regime}.

\subsection{Towards inflation}\label{sec:Towards_Inflation}

\begin{figure}
\centering
 \includegraphics[width=0.9\textwidth]{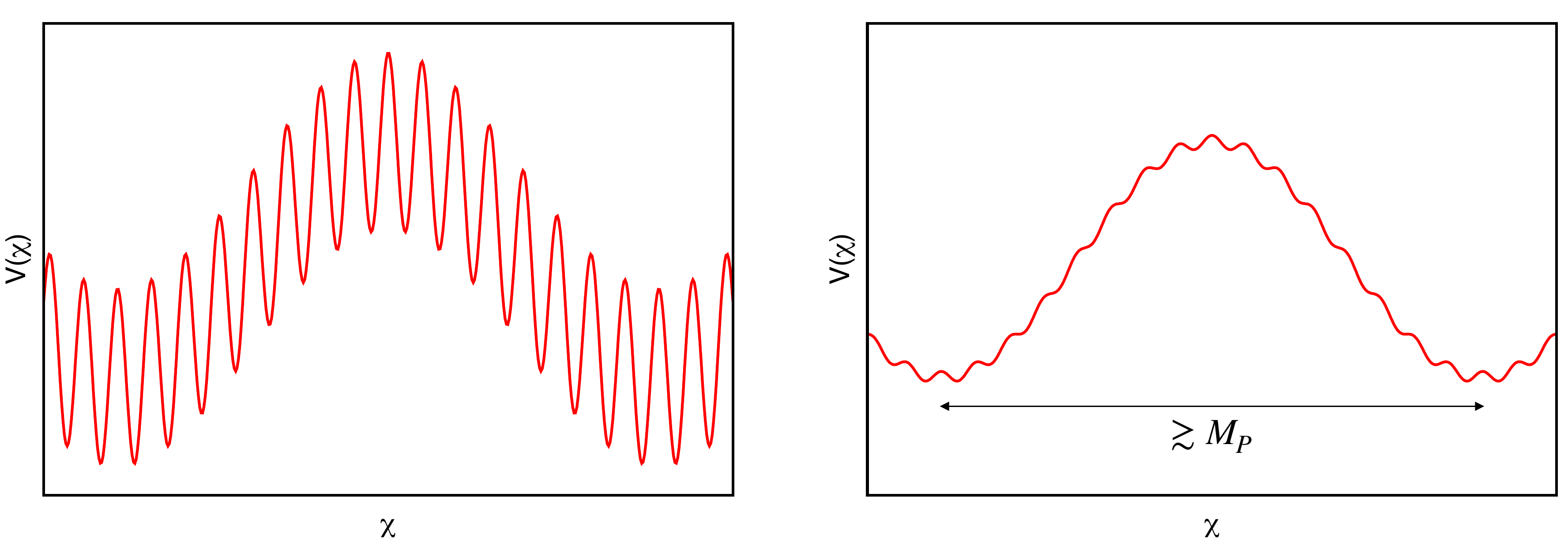}
\caption{Examples of axion potentials with a large field range. The potential on the right-hand side has a large monotonic region as required by inflation.}\label{fig:pot}
\end{figure}

Let us finally come to the issue of realising inflation using the axion $\chi$. While inflationary model building is not the focus of this paper, we stress that it is more difficult than just constructing an EFT for an axion with a large field range. The reason is that we then have to fulfill further constraints in addition to \eqref{regime}. In particular, in order to have a large monotonic region in the inflaton potential, the instanton of large periodicity is required to dominate over the short-period instantons in the potential \eqref{chi-pot} (see Fig.~\ref{fig:pot}), i.e.,
\begin{equation}\label{eq:LSH}
m_{\chi}^\text{long}\gg m_{\chi}^\text{short}\, .
\end{equation}
The issue of higher harmonics has also been analysed in the context of KNP \cite{Kim:2004rp} and, if the effect is not too strong, this can be valuable for phenomenology \cite{Kappl:2015esy}.

Recall that, according to the discussion in Sect.~\ref{sec:ThreeAxions}, the required hierarchy between $m_{\chi}^\text{long}$ and $ m_{\chi}^\text{short}$ can plausibly be realised if we consider an alignment of three (or more) axions.
For instance, in the setup of Fig.~\ref{fig:3}, the light axion is aligned with the $a_1$-$a_2$-plane in the axion field space, and there is no apparent obstruction to stabilising the moduli such that $\mathrm{e}^{-u_\text{short}} \text{``=''}\lbrace\mathrm{e}^{-u_1},\,\mathrm{e}^{-u_2}\rbrace\ll \mathrm{e}^{-u_3}=\mathrm{e}^{-u_\text{long}}$. With \eqref{eq:ShortLongMass}, this then indeed implies \eqref{eq:LSH}.

In addition, there are phenomenological constraints. First, in order to obtain a positive inflaton potential at the end of inflation, we require a suitable uplift of the LVS AdS minimum. This is a non-trivial step since an uplift term in the scalar potential depends on the moduli and can therefore destroy the delicate stabilisation scheme worked out in Sect.~\ref{sec:Corrections}.

Second, large-field inflation requires that the inflaton mass $m_{\chi}^\text{long}$ is of the order $10^{13}$GeV. Since $m_{\tau_{b}}\gg m_{\chi}^\text{long}$, we then have
\begin{equation}\label{eq:eqeq} 
\dfrac{|\cW_{0}|}{g_{s}^{1/4}\sqrt{\cA}\cV^{3/2}} \gg 10^{-5}
\end{equation}
in Planck units, where we used $\mathrm{e}^{\cK_\text{cs}}=1/\cA$.
In addition, we have the constraint that the gravitino mass should be smaller than the Kaluza-Klein scale \cite{31}. This yields \cite{Hebecker:2014kva, Buchmuller:2015oma, Hebecker:2015rya}
\begin{equation}
\dfrac{m_{3/2}}{m_\text{KK}} \sim \frac{\sqrt{g_s}|\W_0|}{\cA^{1/3}\cV^{1/3}} \ll 1\, .
\end{equation}
In general, this is not an issue since $\cV$ is exponentially large (cf.\ Sects.~\ref{sec:mass-scales}, \ref{sec:Corrections}). However, together with \eqref{eq:eqeq}, we obtain the tight constraint
\begin{equation}
10^{-5} \sqrt{\cA} \cV^{3/2} g_{s}^{1/4} \ll |\W_0| \ll \frac{\cA^{1/3}\cV^{1/3}}{\sqrt{g_s}}\,, \label{vconstr}
\end{equation}
which is in tension with the requirement of exponentially large $\cV$. In particular, \eqref{vconstr} implies (forgetting about a factor $\cA^{1/6}$)
\begin{equation}
\cV^{7/6}g_s^{3/4} \ll 10^5\,. \label{vconstr2}
\end{equation}
At the LVS minimum, the volume satisfies
\begin{equation}
\cV \sim \exp(a_s\tau_s) \sim \exp\left(\frac{\alpha}{g_s} \right)\,,\quad  \alpha\approx 0.07 \left(\frac{-\chi}{c}\right)^{2/3}\,,
\end{equation}
where we used $a_s=2\pi$, $\tau_s = \frac{\xi^{2/3}}{(2c)^{2/3}g_s}$ and $\xi=-\frac{\chi\zeta(3)}{2(2\pi)^3}$. Here, $\chi$ is the Euler characteristic, which is negative in the LVS, $\chi\le -2$. The number $c$ is related to the triple self-intersection number of the cycle associated to $\tau_s$ and can be shown to obey $c \le \frac{\sqrt{2}}{3}$ on any Calabi-Yau with a blow-up cycle of the type necessary for the LVS \cite{35}. For the simple example $\mathbb{CP}^4_{(1,1,1,6,9)}[18]$ discussed in this section, we have $\chi=-540$ and $c=\frac{1}{9\sqrt{2}}\approx 0.079$ such that $\alpha\approx 25.8$. The volume is therefore extremely large even at the boundary of perturbative control, i.e., $\cV \gtrsim \mathcal{O}(10^{11})$ for $g_s\lesssim 1$. It is therefore impossible to satisfy \eqref{vconstr2} on this particular manifold. However, the situation may improve on other Calabi-Yaus with a smaller $|\chi|$ and/or a larger $c$. In particular, because of $\chi\le -2$ and $c \le \frac{\sqrt{2}}{3}$, we have $\alpha \gtrsim 0.19$. For values of $\alpha$ sufficiently close to this bound, there is a small window in which inflation might be realisable. We leave a detailed study of the winding scenario on such manifolds for future work.

\section{Conclusions}
\label{sec:summary}

In this paper, we studied winding trajectories of complex-structure axions in type IIB flux compactifications (see Fig.~\ref{fig:almal}). We argued that large-$f$ effective axions can be constructed along such trajectories and discussed several concrete proposals to realise this idea. (For recent results in a different approach see \cite{Hebecker:2018yxs, Buratti:2018xjt}.) 

We first studied the simplest setting of aligned winding in a $2$-axion field space, where the effective axion is aligned with one of the two fundamental axions.
We found a general no-go theorem ruling out large $f$ in such models on any Calabi-Yau. Our result is based on the observation that the flux choice required for the alignment leads to a hierarchy in the saxion vevs. This hierarchy has the effect of cancelling a naive enhancement factor in $f$ and thereby constrains $f$ to be sub-Planckian.

We offered three alternatives to circumvent this issue. First, we discussed the idea of misaligned winding, where the effective axion is aligned with (for
example) a diagonal in the $2$-axion plane. We found that a hierarchy between the saxion vevs is avoided in this setting such that parametrically large $f$ appears possible. Such models can be shown to not admit large monotonic regions in the axion potential and are therefore not suitable for large-field inflation. They are, however, interesting as potential examples of large field ranges in string theory.

In a second proposal, we included exponentially suppressed corrections in the superpotential. Such corrections can soften the dangerous hierarchy in the saxion vevs if the superpotential is fine-tuned to be very small. This is interesting because both large field ranges and large monotonic regions in the potential may be realised this way.

Third, we considered an extension of the winding proposal to three or more fundamental axions. In particular, we showed that, contrary to the 2-axion case, aligned winding in a 3-axion field space does not necessarily lead to a large hierarchy between the saxion vevs. This suggests that such models can realise large field ranges and large monotonic regions in the axion potential. Developing this idea further, we performed a general study of (mis-)aligned winding in the large-complex-structure limit, where we allowed the winding trajectory to be a combination of an arbitrary number of axions. If the axion number is at least 4 and certain geometric conditions on the mirror-dual triple intersection numbers are satisfied, some of the flux parameters do not backreact on the saxion vevs. These fluxes can potentially be utilised to engineer a long winding trajectory. This idea could be a promising starting point for constructing explicit alignment models in type IIB with super-Planckian monotonic regions in the axion potential. It would be very interesting to study this and the other proposals discussed above further on explicit Calabi-Yaus.

Our results may have important implications for axionic versions of the WGC and other swampland conjectures. In particular, the Smallest Charge WGC requires $f$ to be sub-Planckian in the regime of a controlled instanton expansion. Finding explicit geometries that realise (mis-)aligned winding would therefore imply a parametric violation of the Smallest Charge WGC. From the perspective of the related (Sub-)Lattice WGC, this would correspond to a parametrically large coarseness of the sub-lattice populated by the WGC states. Furthermore, models of aligned winding with super-Planckian monotonic regions in the axion potential would imply a parametric violation of the Strong WGC. Finally, our constructions can be argued to parametrically violate the refined version of the Swampland Distance Conjecture.

With the aim of building a low-energy EFT for the large-$f$ effective axion, we reconsidered the LVS in the presence of a light complex-structure axion. Our discussion of the relevant mass scales suggests that there are no general obstructions to constructing such an EFT. In particular, we computed possible corrections spoiling the large field range and showed that they are suppressed for large enough volume and very small string coupling $g_{s}$. The requirement of very small $g_s$ is relaxed if one assumes that the complex-structure dependence of non-perturbative corrections to the superpotential is negligible. We gave an argument for why this is indeed expected if the cycle supporting the non-perturbative effect is a local blow-up. It might be interesting to understand, also independently of our initial motivation, non-perturbative effects on these local blow-ups in more details.

One of the primary motivations for studying super-Planckian field ranges in string theory is large-field inflation. We analysed whether such models can be realised in the context of the winding scenario. Our conclusion is rather negative: we found strong constraints that are in tension with the exponentially large volume in the LVS. Nevertheless, we also found a small window where large-field inflation might be realisable and gave concrete bounds on the required topological data of candidate Calabi-Yaus. It would be interesting to study this possibility further in future work.
\\

{\bf Acknowledgments}

We thank Eran Palti for initial collaboration and many useful discussions. We would also like to thank Christoph Mayrhofer and Thomas Van Riet for helpful correspondence.

\bibliographystyle{jhep}
\bibliography{refs-swamp}

\end{document}